\documentclass[superscriptaddress, aps, preprintnumbers,prd,
amsmath, amssymb, sort&compress, nofootinbib, 10pt, paper, floatfix]{revtex4-2}

\usepackage{fnpct} 
\usepackage{comment}
\usepackage{ifpdf}
\usepackage{slashed}
\usepackage{color}
\usepackage[mathscr]{eucal}
\usepackage[utf8]{inputenc}
\usepackage{cancel}

\usepackage{braket}

\usepackage[most]{tcolorbox}

\tcbset{colback=white, colframe=black, 
        highlight math style= {enhanced, %<-- needed for the ’remember’ options
            colframe=red,colback=red!10!white,boxsep=0pt}
        }

\ifpdf
  \usepackage{graphicx}     %   usepackage without driver option
  \usepackage[bookmarksopen,colorlinks=true,linkcolor=bblue,citecolor=bblue,urlcolor=ppink]{hyperref}
\else     % For (p)LaTeX + dvipdfmx
\fi

\definecolor{red}{rgb}{1,0,0}
\definecolor{darkred}{rgb}{0.6,0,0}
\definecolor{darkgreen}{rgb}{0.992447,0.623778,0.034597}
\definecolor{ppink}{rgb}{1,0.4,0.4}
\definecolor{bblue}{rgb}{0.284602,0.317763,0.963947}
\definecolor{purple}{rgb}{0.5 ,0, 0.7}

\definecolor{dgreen}{rgb}{0 ,0.5, 0.5}

\newcommand{\vev}[1]{ \left< {#1} \right> }
\newcommand{\dd}{\mathrm{d}}

\newcommand{\ee}{\text{e}}

\newcommand{\tmax}{\text{max}}
\newcommand{\tmin}{\text{min}}

\makeatletter
\newcommand\footnoteref[1]{\protected@xdef\@thefnmark{\ref{#1}}\@footnotemark}
\makeatother

\allowdisplaybreaks[1]

\begin{document}

\title{
Overlap reduction functions for pulsar timing arrays and astrometry
}

\author{Keisuke Inomata}
\email{inomata@jhu.edu}
\affiliation{
 William H.\ Miller III Department of Physics \& Astronomy, Johns Hopkins University, 3400 N.\ Charles St., Baltimore, MD 21218, USA
}

\author{Marc Kamionkowski}
\email{kamion@jhu.edu}
\affiliation{
 William H.\ Miller III Department of Physics \& Astronomy, Johns Hopkins University, 3400 N.\ Charles St., Baltimore, MD 21218, USA
}

\author{Celia M.\ Toral}
\email{celia.m.toral@Vanderbilt.Edu}
\affiliation{Department of Physics and Astronomy, Vanderbilt University, 2301 Vanderbilt Place, Nashville, Tennessee 37235, USA}

\author{Stephen R.\ Taylor}
\email{stephen.r.taylor@Vanderbilt.Edu}
\affiliation{Department of Physics and Astronomy, Vanderbilt University, 2301 Vanderbilt Place, Nashville, Tennessee 37235, USA}

\begin{abstract}
We present an efficient technique for calculating the angular two-point correlation functions (or ``overlap reduction functions'') induced by gravitational waves in both the pulse arrival times of pulsars and in the angular deflections of distant sources.  In the most general case, there are six auto- and cross-correlations for the pulse arrival times and the two components of the angular deflection. We provide results for spin-2 (i.e., general-relativistic) gravitational waves as well as the spin-1 modes that may arise in alternative-gravity theories. These calculations can be easily implemented for future analysis or study, and we provide code to do so.
\end{abstract}

\maketitle

\section{Introduction}

Evidence of a stochastic gravitational-wave background (SGWB) with periods of years to decades has recently been presented by four pulsar-timing array (PTAs) collaborations~\cite{NANOGrav:2023gor,EPTA:2023fyk,Reardon:2023gzh,Xu:2023wog}, broadening the landscape of observational gravitational wave (GW) science to encompass almost 11 orders of magnitude in frequency. These new observations call for increased efforts to characterize GWs at these $\sim$nHz frequencies.  Some obvious next milestones beyond this first evidence include the search for anisotropy and polarization in the SGWB.  Detection of either would provide additional information on the sources of these GWs which, at these frequencies, are expected to be from a population of binary supermassive black holes at mass scales of $\sim10^8-10^{10}M_\odot$~\cite{Rajagopal:1994zj,Jaffe:2002rt,Sesana:2008mz,Phinney:2001di,Middleton:2015oda,Sato-Polito:2023spo,Gardiner:2023zzr,Sato-Polito:2023gym}, 
but may also arise from other, more exotic, sources
\cite{Olmez:2010bi, Sousa:2013aaa, Miyamoto:2012ck, Kuroyanagi:2012jf, Caprini:2010xv, Starobinsky:1979ty, Zhao:2013bba, Inomata:2016rbd, NANOGrav:2023hvm}.

The measurement process for PTAs relies on the apparent change to the spin period of pulsars induced by the passage of a GW between a pulsar and the Earth. The integrated effect over the path of photons from the pulsar to the Earth leads to a redshift to the pulse arrival rate, and (upon a further time integral) a deviation to the expected arrival time of radio pulses. By contrast, a SGWB at frequency ranges that overlap with the PTA sensitivity band may also be sought with astrometry~\cite{Braginsky:1989pv,Kaiser:1996wk,Book:2010pf,Qin:2018yhy,Caliskan:2023cqm}. The corresponding effect here is to induce deflections to measured positions of sources, such as stars or quasars. As such, one may consider PTAs and astrometry together to probe GWs through the $4$-vector of light deflections. The key SGWB observable in astrometry is the angular two-point correlation function~\cite{Book:2010pf} of the positional deflections of widely separated sources, an analog of the Hellings-Downs curve~\cite{Hellings:1983fr} for pulsar time delays due to a Gaussian, stationary, isotropic, and unpolarized SGWB. Using astrometry, GAIA and extragalactic radio sources constrain the energy density of the
SGWB to be $< 1.1\%$ of the Universe's critical density for frequencies $6\times 10^{-18}~\mathrm{Hz} \lesssim f \lesssim 10^{-9}~\mathrm{Hz}$~\cite{Darling:2018hmc}, which is quite a bit weaker than the values $\sim10^{-8}$ implied by recent PTA results~\cite[e.g.,][]{NANOGrav:2023gor}.  However, there are prospects for future missions that might provide improved astrometric measurements~\cite{Boehm:2017wie,Fedderke:2022kxq}, and such measurements may also be possible with photometric surveys \cite{Wang:2020pmf,Wang:2022sxn,Pardo:2023cag}.

If the SGWB is isotropic and unpolarized---as most searches assume---then the two-point correlation functions depend only on the angle between the two points of measurement.  If, however, the SGWB is anisotropic or polarized, then there may be more structure in the two-point correlation, analogous to that previously computed in a PTA context~\cite{Mingarelli:2013dsa,Gair:2014rwa,Taylor:2013esa,NANOGrav:2023tcn,Chu:2021krj,Liu:2022skj,AnilKumar:2023kvt,Kato:2015bye,Sato-Polito:2021efu,AnilKumar:2023yfw,AnilKumar:2023hza,Bernardo:2023jhs}. Indeed, the generalization of the astrometric angular two-point correlation function has already been performed for the case of a SGWB composed of non-Einsteinian GW polarizations~\cite{Mihaylov:2018uqm,OBeirne:2018slh}, which
echoes analogous work for pulsar timing~\cite{LeeJenetPrice:2008,Chamberlin:2011ev,Gair:2015hra,Qin:2018yhy,AnilKumar:2023yfw,Bernardo:2023jhs}. Here, we present the first calculation of angular two-point correlations induced in astrometric deflections by anisotropies and linear or circular polarization in the SGWB.  We generalize here the approach for timing residuals of Ref.~\cite{AnilKumar:2023yfw} to the present case of angular deflections.  This allows for fairly compact expressions for the correlation functions that are easily evaluated numerically.  Results are presented for the spin-2 (tensor) GW polarizations that arise in general relativity, as well as the spin-1 (vector) polarizations that may arise in alternative-gravity theories~\cite{Hellings:1973zz,BeltranJimenez:2009lpn}. 
Similar to the previous work~\cite{AnilKumar:2023yfw}, we will not study the case of spin-0 (scalar) GWs, decomposed into scalar-transverse (ST) and scalar-longitudinal (SL) modes, for the following reasons.
For ST modes, the redshift response and the deflection angle consist only of the monopole and the dipole moments~\cite{Qin:2018yhy}, which limits the information of anisotropies of GW background that we can extract from the observations. 
For SL modes, we need a more careful treatment than for the spin-2 or spin-1 cases because the redshift response depends on the pulsar term~\cite{Cornish:2017oic,Qin:2018yhy}, whose analysis is beyond the scope of this paper.

We summarize here for the busy reader the key equations and relationships that can be used right away. First, we note that the deviation to a pulse arrival time is a scalar function on the sphere, while the angular deflection of a point's measured position is a vector on the two-dimensional celestial sphere.  There are, therefore, most generally 6 possible auto- and cross-correlations for the combined data set from a PTA and an astrometry survey, all of which we derive.  We also present details of the derivations, and as a result, the paper is long.  Still, the central results are fairly compact.  The most general PTA overlap reduction functions (ORFs), reprised from Ref.~\cite{AnilKumar:2023yfw}, can be obtained from Eq.~(\ref{eq:orfzz}) with coefficients $F^L_{\ell\ell'}$ given in Table~\ref{tab:F_t_zz} (for intensity anisotropy and circular and linear polarizations for spin-2 and spin-1 GWs).  The ORFs for angular deflections are given in Eq.~(\ref{eq:orf}) with coefficients listed in Tables~\ref{tab:F_tb}\,-\,\ref{tab:F_bb}, which are derived here for the first time. For the cross-correlations between the PTA and the astrometry signals (also derived here for the first time), the ORFs are given in Eqs.~(\ref{eq:orf_zeb}) and (\ref{eq:orf_ebz}) with the coefficients given in Eq.~(\ref{eq:coeff_cross}).

This paper is laid out as follows. In the following Section, we describe the characterization of the SGWB in terms of spherical-harmonic expansion coefficients for the intensity and circular- and linear-polarization patterns.  Section \ref{sec:orfs} then calculates the ORFs, starting first with PTA timing residuals and then moving on to astrometry observables, and in each case, beginning with the spin-2 calculations and then following up with the spin-1 calculations.  
We make closing remarks in Section \ref{sec:conclusions}, and an Appendix provides some relevant properties of Wigner D matrices. 
We note that the code for the numerical calculation of the ORFs is available at \href{https://github.com/KeisukeInomata0/pyORFs}{our Github page}.

\section{Characterization of the gravitational-wave background}
\label{sec:characterization}

In this work, we consider both spin-2 (tensor) and spin-1 (vector) GWs.
We take the same characterization for the spin-2 GWs as Ref.~\cite{AnilKumar:2023kvt} and generalize it to the spin-1 GWs with the convention in Ref.~\cite{Jeong:2012df}.
We expand the tensor fluctuation as 
\begin{align}
  h_{ij}(t,\bm x) &= \int \dd f \int \dd^2 \hat \Omega \left[\sum_{\lambda = +,\times,x,y} \tilde h_\lambda(f,\hat \Omega) \ee^\lambda_{ij}(\hat \Omega) \right] \ee^{-2\pi i f(t - \hat \Omega \cdot \bm x)},
\end{align}
where $\hat \Omega$ is the unit vector for the propagation direction of the GWs.
The explicit forms of the polarization tensors are given by 
\begin{align}
  \ee^+_{ij}(\hat \Omega) &= \hat p_i \hat p_j- \hat q_i \hat q_j, \ \ee^\times_{ij}(\hat \Omega) = \hat p_i \hat q_j + \hat q_i \hat p_j, \\
  \ee^x_{ij}(\hat \Omega) &= \hat \Omega_i \hat p_j + \Omega_j \hat p_i, \ \ee^y_{ij}(\hat \Omega) = \hat \Omega_i \hat q_j + \hat \Omega_j \hat q_i,
\end{align}
where $\hat p$ and $\hat q$ are the unit vectors perpendicular to $\hat \Omega$ and they satisfy $\hat p \cdot \hat q = 0$.
The polarizations in the first (second) line are for the spin-2 (spin-1) GWs.
If we set $\hat p$, $\hat q$, and $\hat \Omega$ to be aligned with the $x$, $y$, and $z$ axes respectively, the nonzero components of each polarization vector are $\epsilon^+_{xx} = - \epsilon^+_{yy} = 1$, $\epsilon^\times_{xy} = \epsilon^\times_{yx} = 1$, $\epsilon^x_{xz}= \epsilon^x_{zx} = 1$, and $\epsilon^y_{yz}= \epsilon^y_{zy} = 1$.
Note that we take the normalization that gives $\epsilon^{\lambda\,ij} \epsilon^{\lambda'}_{ij} = 2 \delta^{\lambda \lambda'}$.
We hereafter assume that the spin-2 and the spin-1 GWs are uncorrelated. 
We formulate the anisotropies of the spin-2 and the spin-1 GWs, respectively, through
\begin{align}
  \vev{\tilde h_{\lambda^t}^*(f,\hat \Omega)\tilde h_{{\lambda^t}'}(f',\hat \Omega')} = \delta_D(f-f') \delta_D(\hat \Omega, \hat \Omega') \mathcal P^t_{\lambda^t {\lambda^t}'}(f, \hat \Omega), \\
  \vev{\tilde h_{\lambda^v}^*(f,\hat \Omega)\tilde h_{{\lambda^v}'}(f',\hat \Omega')} = \delta_D(f-f') \delta_D(\hat \Omega, \hat \Omega') \mathcal P^v_{\lambda^v {\lambda^v}'}(f, \hat \Omega),
\end{align}
where $\lambda^t \in \{+,\times\}$ and $\lambda^v \in \{x,y\}$, $\delta_D$ is the delta function, and $\mathcal P_{\lambda \lambda'}$ is the spectral density of the GW background, which depends on the frequency, the polarization, and the propagation direction of GWs. 
We parametrize the spectral density as 
\begin{align}
  \mathcal P^{p}_{\lambda \lambda'} =\begin{pmatrix}
  I^p(f,\hat\Omega) + Q^p(f,\hat\Omega) & U^p(f,\hat\Omega) - i V^p(f,\hat\Omega) \\
  U^p(f,\hat\Omega) + iV^p(f,\hat\Omega) & I^p(f,\hat\Omega)- Q^p(f,\hat\Omega)\\ 
  \end{pmatrix},
\end{align}
where $p \in \{t,v\}$.
Specifically, we obtain 
\begin{align}
  I^t(f,\hat \Omega) &= \frac{1}{2} \vev{|\tilde h_+|^2 + |\tilde h_\times|^2} = \frac{1}{2} \vev{ |\tilde h^t_L|^2 + |\tilde h^t_R|^2}, \ 
  Q^t(f,\hat \Omega) = \frac{1}{2} \vev{|\tilde h_+|^2 - |\tilde h_\times|^2}, \nonumber \\
  U^t(f,\hat \Omega) &= \frac{1}{2} \vev{ \tilde h^{\,*}_+ \tilde h_\times + \tilde h^*_{\times} \tilde h_+}, \ 
  V^t(f,\hat \Omega) = \frac{1}{2i} \vev{ \tilde h^{*}_+ \tilde h_\times - \tilde h^*_{\times} \tilde h_+} = \frac{1}{2} \vev{ |\tilde h^t_L|^2 - |\tilde h^t_R|^2}, \\
  I^v(f,\hat \Omega) &= \frac{1}{2} \vev{|\tilde h_x|^2 + |\tilde h_y|^2} = \frac{1}{2} \vev{ |\tilde h^v_L|^2 + |\tilde h^v_R|^2}, \ 
  Q^v(f,\hat \Omega) = \frac{1}{2} \vev{|\tilde h_x|^2 - |\tilde h_y|^2}, \nonumber \\
  U^v(f,\hat \Omega) &= \frac{1}{2} \vev{ \tilde h^{*}_x \tilde h_y + \tilde h^*_{y} \tilde h_x}, \ 
  V^v(f,\hat \Omega) = \frac{1}{2i} \vev{ \tilde h^*_{x} \tilde h_y - \tilde h^*_{y} \tilde h_x} = \frac{1}{2} \vev{ |\tilde h^v_L|^2 - |\tilde h^v_R|^2},
\end{align}
where the arguments of $\tilde h$ have been omitted and the left-handed and the right-handed modes are defined as $\tilde h^t_L = (\tilde h_+ - i \tilde h_\times)/\sqrt{2}$, $\tilde h^t_R = (h_+ + i h_\times)/\sqrt{2}$, $\tilde h^v_L = (\tilde h_x - i \tilde h_y)/\sqrt{2}$, and $\tilde h^v_R = (\tilde h_x + i \tilde h_y)/\sqrt{2}$.
For convenience, we define $P^\pm$ as 
\begin{align}
  P^{t, \pm}(f,\hat \Omega) & \equiv Q^t \pm i U^t = \frac{1}{2} \vev{ (\tilde h_+ \mp i \tilde h_\times)^* (\tilde h_+ \pm i \tilde h_\times)}, \\
  P^{v, \pm}(f,\hat \Omega) & \equiv Q^v \pm i U^v = \frac{1}{2} \vev{ (\tilde h_x \mp i \tilde h_y)^* (\tilde h_x \pm i \tilde h_y)}, 
\end{align}
and can rewrite this as 
\begin{align}
  P^{p,+}(f,\hat \Omega) = \vev{\tilde h^{p*}_{L} \tilde h^p_R}, \ 
  P^{p,-}(f,\hat \Omega) = \vev{\tilde h^{p*}_R \tilde h^p_L}.
\end{align}
These Stokes parameters are expanded as 
\begin{align}
  \label{eq:gw_i}
  I^p(f,\hat\Omega) &= I^p_0(f) \sum^\infty_{L=0} \sum^L_{M = -L} c^{p,I}_{LM}(f) Y_{LM}(\hat \Omega), \\
  V^p(f,\hat\Omega) &= I^p_0(f) \sum^\infty_{L=0} \sum^L_{M = -L} c^{p,V}_{LM}(f) Y_{LM}(\hat \Omega), \\ 
  \label{eq:gw_p} 
  P^{t,\pm}(f,\hat\Omega) &= I^t_0(f) \sum^\infty_{L=4} \sum^L_{M = -L} c^{t,\pm}_{LM}(f) \,_{\pm 4}Y_{LM}(\hat \Omega), \\
  \label{eq:gw_m}
  P^{v,\pm}(f,\hat\Omega) &= I^v_0(f) \sum^\infty_{L=2} \sum^L_{M = -L} c^{v,\pm}_{LM}(f) \,_{\pm 2}Y_{LM}(\hat \Omega),
\end{align}
where we normalize $I^p_0(f)$ such that $c^{p,I}_{00}(f) = 1$.
For later convenience, we here introduce the spherical-harmonic coefficients for the scalar $E$ and pseudo-scalar $B$ modes in the linear polarization map ($P^\pm$), analogous to the CMB polarizations:
\begin{align}
   c^{p,E}_{LM}(f) = \frac{1}{2}( c^{p,+}_{LM}(f) + c^{p,-}_{LM}(f)), \    c^{p,B}_{LM}(f) = \frac{1}{2i}( c^{p,+}_{LM}(f) - c^{p,-}_{LM}(f)).
\end{align}

\section{Overlap reduction functions}
\label{sec:orfs}

We consider two quantities. 
One is the redshift response $z$ of PTA observation for GWs, and the other is a deflection-angle field $\delta \vec n$ in astrometry probes.
We will obtain the exact formulas for all the ORFs associated with PTA and astrometry, while showing the numerical results for some of them. 
We refer to \href{https://github.com/KeisukeInomata0/pyORFs}{our Github page} for the numerical results of all the ORFs.

\subsection{Auto-correlation of redshift responses}
We first consider the auto-correlation of the redshift responses in PTA observations. 
The redshift response is related to the GWs propagating along $\hat \Omega$ as~\cite{Maggiore:2018sht}
\begin{align}
  z(t,\hat n_a,\hat \Omega) = \frac{\hat n_a^i \hat n_a^j \Delta h_{ij}(t,\hat \Omega)}{2(1+\hat \Omega \cdot \hat n_a)},
  \label{eq:z_nk}
\end{align}
where the subscript $a$ labels a given pulsar and $L_a \hat n_a$ is the position of the pulsar from solar system barycenter, located at $\vec 0$.
We here define $\Delta h_{ij} \equiv h_{ij}(t, \hat \Omega, \vec 0) - h_{ij}(t_p, \hat \Omega, L_a \hat n_a)$ as the difference of the tensor fluctuations between the solar barycenter at $t$ and the pulsar at $t_p = t-L_a$.
However, similar to existing analysis, such as Ref.~\cite{AnilKumar:2023yfw}, we disregard the contributions of the tensor fluctuations at the pulsar throughout this work.

Although the ORF for this case was studied in Ref.~\cite{AnilKumar:2023yfw}, we summarize the calculation for completeness. 
Throughout this work, we focus on the contribution from a single frequency bin $f$.
Then, we can expand $z$ as 
\begin{align}
  z_f(\hat n) &= \sum_{\ell m} Y_{\ell m}(\theta, \phi) z_{f,\ell m}.
\end{align}
Note that $z_f$ is a Fourier mode and can be complex and the subscript $f$ is the shorthand notation for the Fourier mode argument, e.g. $z_f = z(f)$, throughout this work.

We consider the two-point correlation function:
\begin{align}
  \vev{z^*_f(\hat n_a) z_f(\hat n_b)} &= \sum_{\ell m} \sum_{\ell' m'}\vev{z^*_{f,\ell m} z_{f,\ell' m'}} Y^*_{\ell m}(\hat n_a) Y_{\ell' m'}(\hat n_b) \nonumber \\
  &= \sum_{p=t,v} \sum_X i^X \left[ I^p_0(f) \sum_{L,M} c^{p,X}_{LM}(f) \Gamma^{p,X,zz}_{LM}(\hat n_a, \hat n_b) \right],
  \label{eq:zz_corr}
\end{align}
where the average is over the ensemble of realizations of the GW background, and $X \in \{I,V,E,B\}$ and $i^X = i$ for $X=B$ and $i^X=1$ for the others throughout this work.\footnote{Note that the convention for $i^X$ for $V$ differs from that in Ref.~\protect\cite{AnilKumar:2023yfw}.}\ Here, $\Gamma_{LM}$ is the ORF, given by 
\begin{tcolorbox}[ams equation]
 \Gamma^{p,X,zz}_{LM}(\hat n_a, \hat n_b) = (-1)^L \sqrt{\pi} \sum^{\ell_\tmax}_{\ell=\ell_\tmin} \sum^{\ell_\tmax}_{\ell' = \ell_\tmin} F^{L,p,X,zz}_{\ell \ell'}\left\{ Y_\ell(\hat n_a) \otimes Y_{\ell'}(\hat n_b)\right\}_{LM},
  \label{eq:orfzz}
\end{tcolorbox}  
where 
\begin{align}
  \left\{ Y_{\ell}(\hat n_a) \otimes Y_{\ell'}(\hat n_b)\right\}_{LM} = \sum^{\ell}_{m=-\ell} \sum^{\ell'}_{m'=-\ell'}\Braket{\ell m \ell' m' |L M} Y_{\ell m}(\hat n_a)Y_{\ell' m'}(\hat n_b)
\end{align}
is a bipolar spherical harmonic (BiPoSH)~\cite{Hajian:2003qq,Hajian:2005jh,Joshi:2009mj,Book:2011na} and we have explicitly shown the lower and the upper bounds for the sums.
We will see that $\ell_\tmin = 2$ for the spin-2 GWs and $\ell_\tmin = 1$ for the spin-1 GWs below Eqs.~(\ref{eq:z_t}) and (\ref{eq:z_vl}).
Strictly speaking, $\ell_\tmax = \infty$. However, the numerical calculation of the ORFs with $\ell_\tmax = \infty$ requires an infinite computational cost.
Fortunately, if we set $\ell_\tmax$ to be large enough, the obtained ORFs become insensitive to a concrete value of $\ell_\tmax$.
In this work, for the numerical calculation, we change the value of $\ell_\tmax$ depending on the ORFs.
In particular, we take larger $\ell_\tmax$ for the ORFs including redshift response due to spin-1 GWs because they slowly converge compared to the other quantities, as we will see below. 
We also note that the E- and the B-mode ORFs ($\Gamma^{E}_{LM}$ and $\Gamma^B_{LM}$) are zero for $L < 4$ for spin-2 GWs and for $L < 2$ for spin-1 GWs because of the properties of the spin-weighted spherical harmonics, $_sY_{\ell m} = 0$ for $\ell < |s|$ (see the lower bounds of $L$ in Eqs.~(\ref{eq:gw_p}) and (\ref{eq:gw_m})).
All the nontrivial information about the ORFs is included in the expressions of $F^{L,p,X,zz}_{\ell \ell'}$, which we will obtain in the following.

\subsubsection{Spin-2 GWs}

We here consider the spin-2 GWs, described by the $+$ and $\times$ modes. 
From Eq.~(\ref{eq:z_nk}), the redshift response due to the tensor perturbations with the frequency $f$ and the propagation direction $\hat k$ becomes
\begin{align}
  z_f(\hat n,\hat k) &= \frac{1}{2} \left[ \tilde h^+(f,\hat k)(1- \cos\theta) \cos(2 \phi) + \tilde h^\times(f,\hat k) (1- \cos\theta) \sin(2 \phi)\right] \nonumber \\
  &= \frac{1}{2\sqrt{2}} \left[ \tilde h^t_L(f,\hat k) (1- \cos\theta) \ee^{2i \phi} + \tilde h^t_R(f,\hat k) (1- \cos\theta) \ee^{-2i\phi}\right],
  \label{eq:z_nk2}
\end{align}
where we have taken the coordinates of $\hat k = (0,0,1)$ and $\hat n = (\sin \theta \cos \phi, \sin\theta \sin \phi, \cos \theta)$.
From this, we can obtain 
\begin{align}
  \tilde z_{f,\ell m}(\hat k) &= \int \dd^2 \hat n \, Y^*_{\ell m}(\hat n) z_f(\hat n,\hat k) \nonumber \\
  & = \frac{z^t_\ell}{\sqrt{2}} \left( \tilde h^t_L(f,\hat k)\delta_{m2} + \tilde h^t_R(f,\hat k) \delta_{m,-2} \right),
  \label{eq:z_lm_t}
\end{align}
where 
\begin{align}
  z^t_\ell \equiv (-1)^\ell \sqrt{\frac{4\pi (2\ell + 1)(\ell-2)!}{(\ell+2)!}}.
  \label{eq:z_t}
\end{align}
We can find $z_{f,\ell m} = 0$ in $\ell < 2$ because $Y_{\ell m} = 0$ for $\ell < |m|$, which leads to $\ell_\tmin = 2$ in Eq.~(\ref{eq:orfzz}) for the spin-2 GWs.
Then, we move to the general case with a propagation direction $\hat \Omega$. 
To this end, we use the following relation with Wigner $D$-matrix, $D^{(\ell)}_{mm'}$~\cite{Khersonskii:1988krb}:
\begin{align}
  z_f(\hat n,\hat \Omega) = \sum_{\ell} \sum_{m m'}Y_{\ell m}(\hat n) D^{(\ell)}_{m m'}(\hat \Omega) \tilde z_{f,\ell m'}(\hat \Omega),
  \label{eq:z_ydz}
\end{align}
where $D^{(\ell)}_{mm'}(\hat \Omega) = D^{(\ell)}_{mm'}(\phi_\Omega,\theta_\Omega,0)$ with $\hat \Omega = (\theta_\Omega, \phi_\Omega)$.
See also Appendix~\ref{app:wigner} for the properties of Wigner $D$-matrix. 
From this, we obtain 
\begin{align}
  \tilde z_{f,\ell m}(\hat \Omega) &= \int \dd \hat n \, Y^*_{\ell m}(\hat n) z_f(\hat n,\hat \Omega) \nonumber \\
  & = \frac{z^t_\ell}{\sqrt{2}} \left(\tilde h^t_{L}(f,\hat \Omega) D^{(\ell)}_{m,2} + \tilde h^t_R(f,\hat \Omega) D^{(\ell)}_{m,-2} \right).
  \label{eq:z_lm_t2}
\end{align}
We here define $z_{f,\ell m} = \int \dd^2 \hat \Omega\, \tilde z_{f,\ell m}(\hat \Omega)$ and obtain 
\begin{align}
  \vev{z^*_{f,\ell m} z_{f,\ell' m'} } &= \frac{z^t_\ell z^t_{\ell'}}{2} \int \dd^2 \hat\Omega \int \dd^2 \hat\Omega' \vev{\left[ \tilde h^{t\,,*}_L(f,\hat \Omega) D^{(\ell)*}_{m,2} + \tilde h^{t\,*}_R(f,\hat \Omega) D^{(\ell)*}_{m,-2} \right] \left[ \tilde h^t_{L}(f,\hat \Omega') D^{(\ell)}_{m,2} + \tilde h^t_R(f,\hat \Omega') D^{(\ell')}_{m',-2} \right]} \nonumber \\
  &= \frac{z^t_\ell z^t_{\ell'}}{2} I^t_0(f) \sum_{LM} \left[ c^{t,I}_{LM}(f) \left( \,^{(0,2,2)}W^{LM}_{\ell \ell' m m'} + \,^{(0,-2,-2)}W^{LM}_{\ell \ell' m m'}\right) + c^{t,V}_{LM}(f) \left(\,^{(0,2,2)}W^{LM}_{\ell \ell' m m'} - \,^{(0,-2,-2)}W^{LM}_{\ell \ell' m m'} \right) \right. \nonumber \\
  & \quad \qquad\qquad\qquad \left.
  + c^{t,+}_{LM}(f) \,^{(4,2,-2)}W^{LM}_{\ell \ell' m m'} + c^{t,-}_{LM}(f) \,^{(-4,-2,2)}W^{LM}_{\ell \ell' m m'} \right].
  \label{eq:zz_w}
\end{align}
$\,^{(abc)}W^{LM}_{\ell \ell' mm'}$ is defined as 
\begin{align}
  \label{eq:W_llmm}
  \,^{(abc)}W^{LM}_{\ell \ell' mm'} &\equiv \int \dd^2 \hat \Omega\, \,_a Y_{LM}(\hat \Omega) D^{(\ell)*}_{mb}(\hat \Omega) D^{(\ell')}_{m'c}(\hat \Omega) \nonumber \\
  &= (-1)^{b+m'}\frac{4\pi}{\sqrt{(2\ell + 1) (2 \ell' + 1)}} \int \dd^2 \hat \Omega\, \,_a Y_{LM}(\hat \Omega) \,_{-b} Y_{\ell m}(\hat \Omega) \,_{c} Y_{\ell' -m'}(\hat \Omega) \nonumber \\
  &= (-1)^{b+m'}\sqrt{4\pi(2L+1)} \begin{pmatrix}
  L & \ell & \ell'\\
  M & m & -m'\\
  \end{pmatrix} 
  \begin{pmatrix}
  L & \ell & \ell'\\
  -a & b & -c\\
  \end{pmatrix} \nonumber \\
  &= 2 (-1)^{2\ell + M + L + b+m'}\sqrt{\pi} \Braket{\ell (-m) \ell' m' |L M} 
  \begin{pmatrix}
  L & \ell & \ell'\\
  -a & b & -c\\
  \end{pmatrix},  
\end{align}
where we have used the following relation between Wigner $D$-matrix and spin-weighted spherical harmonics~\cite{Gair:2015hra},
\begin{align}
  D^\ell_{m' m}(\phi,\theta,\psi) &= (-1)^{m'}\sqrt{\frac{4\pi}{2\ell + 1}} \,_{m}Y_{\ell,-m'}(\theta, \phi) \ee^{-im\psi}, \\\
  [D^\ell_{m' m}(\phi,\theta,\psi)]^* &= (-1)^{m}\sqrt{\frac{4\pi}{2\ell + 1}} \,_{-m}Y_{\ell,m'}(\theta, \phi) \ee^{im\psi},
\end{align}
and the integral of the product of three spherical harmonics~\cite{Khersonskii:1988krb,Gair:2015hra}\footnote{The integral of the product of three Wigner $D$-matrices is explicitly shown in Ref.~\cite{Khersonskii:1988krb}, from which we can obtain Eq.~(\ref{eq:three_pro}).}
\begin{align}
  &\int \dd^2 \hat \Omega \,_{s_1}Y_{\ell_1 m_1}(\hat \Omega) \,_{s_2}Y_{\ell_2 m_2}(\hat \Omega) \,_{s_3}Y_{\ell_3 m_3}(\hat \Omega) \nonumber \\
  &= \sqrt{\frac{(2 \ell_1 + 1)(2\ell_2 + 1)(2\ell_3 + 1)}{4\pi}}
  \begin{pmatrix}
  \ell_1 & \ell_2 & \ell_3\\
  m_1 & m_2 & m_3\\
  \end{pmatrix}
  \begin{pmatrix}
  \ell_1 & \ell_2 & \ell_3\\
  -s_1 & -s_2 & -s_3\\
  \end{pmatrix}.
  \label{eq:three_pro}
\end{align}
Also, Wigner 3-j symbols are related to the Clebsch-Gordan coefficients as
\begin{align}
  \begin{pmatrix}
  L & \ell_1 & \ell_2\\
  M & m_1 & m_2\\
  \end{pmatrix}
  &= \frac{(-1)^{\ell_1 - \ell_2 + M}}{\sqrt{2 L + 1}} \Braket{\ell_1 m_1 \ell_2 m_2 |L (-M)} \nonumber \\
  &= \frac{(-1)^{2 \ell_1 + M + L}}{\sqrt{2 L + 1}} \Braket{\ell_1 (-m_1) \ell_2 (-m_2) |L M}.
\end{align}

Then, we can rewrite Eq.~(\ref{eq:zz_w}) as 
\begin{align}
\vev{z^*_{f,\ell m} z_{f,\ell' m'} } 
  &=  2 z^t_\ell z^t_{\ell'} I^t_0(f) \sum_{LM}  (-1)^{m+L}\sqrt{\pi} \Braket{\ell (-m) \ell' m' |L M} \nonumber \\
  &\quad
  \times\left\{ 
  \left[c^{t,I}_{LM}(f) X^L_{\ell \ell'} - c^{t,V}_{LM}(f) (1-X^L_{\ell \ell'}) \right]
\begin{pmatrix}
  \ell & \ell' & L\\
  -2 & 2 & 0\\
  \end{pmatrix} 
  + 
  \left[c^{t,E}_{LM}(f) X^L_{\ell \ell'} + i c^{t,B}_{LM}(f) (1-X^L_{\ell \ell'}) \right]
\begin{pmatrix}
  \ell & \ell' & L\\
  2 & 2 & -4\\
  \end{pmatrix} 
  \right\},
  \label{eq:z_lm_corr_t}
\end{align}
where $X^L_{\ell \ell'} = 1$ for $\ell + \ell' + L = \text{even}$ and $0$ otherwise.
Substituting this into Eq.~(\ref{eq:zz_corr}), we finally obtain the coefficients $F^L_{\ell \ell'}$ in Eq.~(\ref{eq:orfzz}), summarized in the left column of Table~\ref{tab:F_t_zz}.

\begin{figure}
        \centering \includegraphics[width=0.6\columnwidth]{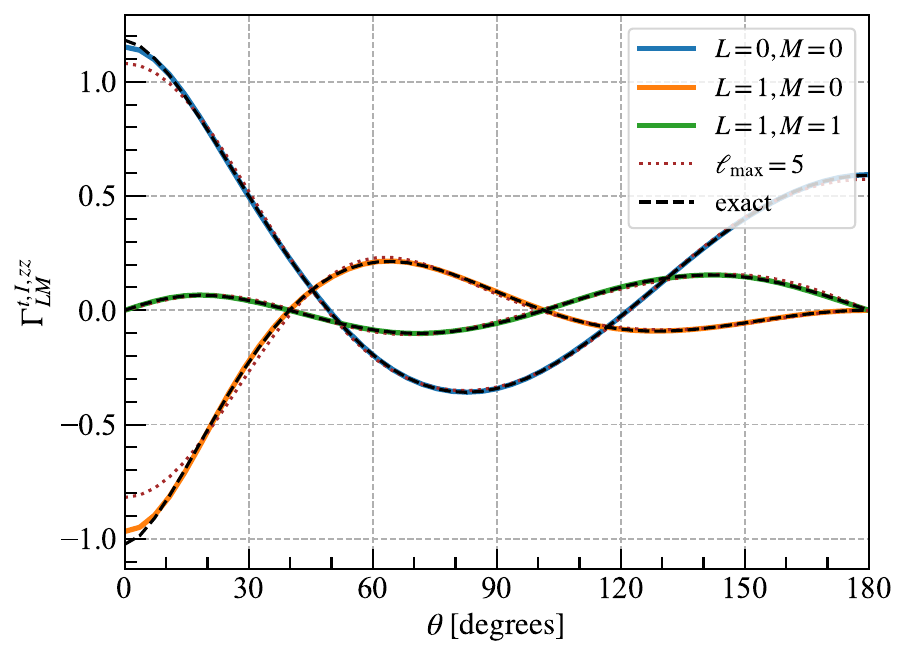}
        \caption{The ORFs of $\Gamma^{t,I,zz}_{LM}(\theta)$ in the coordinates of Eq.~(\ref{eq:coord}).
        The color solid lines are for $\ell_\tmax = 10$.
        The brown dotted lines are for $\ell_\tmax = 5$. 
        The black dashed lines are the exact analytical results obtained in Ref.~\cite{Gair:2014rwa}. 
    }
        \label{fig:g_t_I_zz}
\end{figure}

\begin{figure}
        \centering \includegraphics[width=0.6\columnwidth]{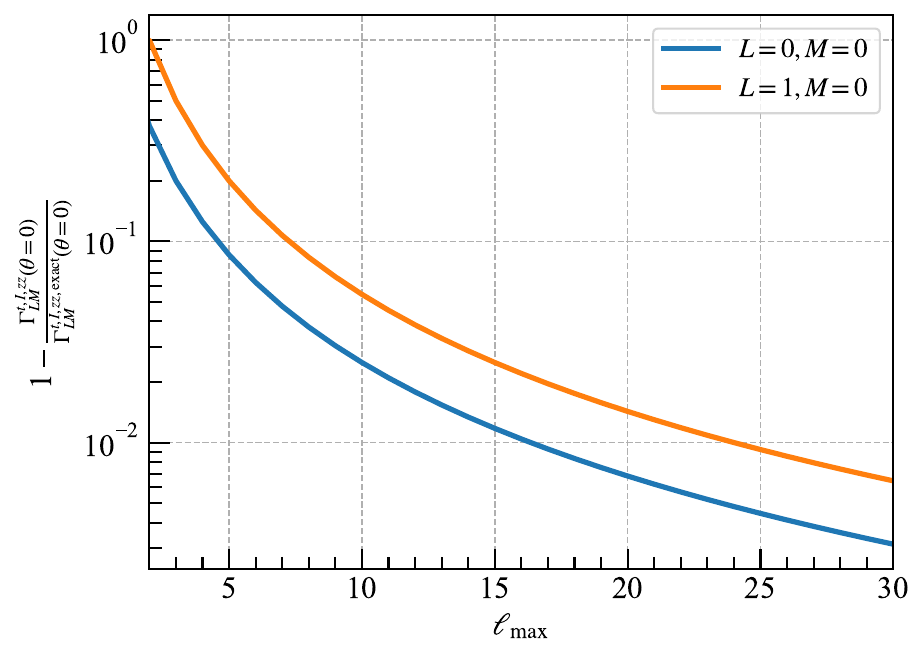}
        \caption{The $\ell_\tmax$ dependence of $\Gamma^{t,I,zz}_{LM}(\theta=0)$ normalized by the exact analytical results obtained in Ref.~\cite{Gair:2014rwa}.
    }
        \label{fig:g_t_I_zz_lmax}
\end{figure}

To get the concrete plots of the ORF, we take the following coordinate choice throughout this work:
\begin{align}
  \hat n_a = (0,0,1),\  \hat n_b = (\sin \theta, 0, \cos \theta).
  \label{eq:coord}
\end{align}
In these coordinates, the ORF only depends on $\theta$.
Figure~\ref{fig:g_t_I_zz} shows the $\theta$ dependence of $\Gamma^{t,I,zz}_{LM}$ with different $\ell_\tmax$, where we also show the exact analytical results in Ref.~\cite{Gair:2014rwa}.
Figure~\ref{fig:g_t_I_zz_lmax} shows the $\ell_\tmax$ dependence of the deviation from the exact value at $\theta=0$, where the deviation becomes the largest.
From these figures, we can see that $\ell_\tmax = 10$ is a good value for the convergence for spin-2 GW case. 
For example, the deviation from the exact value at $\theta =0$ in $L=1$ and $M=0$ is $\sim 5\%$ when $\ell_\tmax =10$.
In general, the value of $\ell_\tmax$ must be tuned to balance the computational time and the required precision for the comparison with real data, while $\mathcal O(1)\,\%$ precision would be sufficient for near-future experiments.

\begin{table*}
\centering
    \everymath{\displaystyle}
    \resizebox{0.8\textwidth}{!}{\begin{tabular}{|c|c|c|}
    \hline
    $X$ & Spin-2 & Spin-1  \\
  \hline\hline

    $I$ & $2 z_\ell^t z_{\ell'}^t  \left( \begin{array}{ccc} \ell & \ell'& L \\ -2 & 2 &  0 \end{array} \right) X^L_{\ell\ell'}$ &  $-2 z_\ell^v z_{\ell'}^v  \left( \begin{array}{ccc} \ell & \ell'& L \\ -1 & 1 &  0 \end{array} \right) X^L_{\ell\ell'}$  \\ \hline
 
    $V$ & $-2 z_\ell^t z_{\ell'}^t  \left( \begin{array}{ccc} \ell & \ell'& L \\ -2 & 2 &  0 \end{array} \right) (1-X^L_{\ell\ell'})$ &  $2 z_\ell^v z_{\ell'}^v  \left( \begin{array}{ccc} \ell & \ell'& L \\ -1 & 1 &  0 \end{array} \right) (1-X^L_{\ell\ell'})$  \\ \hline
    
    $E$ & $2 z_\ell^t z_{\ell'}^t  \left( \begin{array}{ccc} \ell & \ell'& L \\ 2 & 2 &  -4 \end{array} \right) X^L_{\ell\ell'}$ &  $2 z_\ell^v z_{\ell'}^v  \left( \begin{array}{ccc} \ell & \ell'& L \\ 1 & 1 &  -2 \end{array} \right) X^L_{\ell\ell'}$  \\ \hline

    $B$ & $2 z_\ell^t z_{\ell'}^t  \left( \begin{array}{ccc} \ell & \ell'& L \\ 2 & 2 &  -4 \end{array} \right) (1-X^L_{\ell\ell'})$  &  $ 2 z_\ell^v z_{\ell'}^v  \left( \begin{array}{ccc} \ell & \ell'& L \\ 1 & 1 &  -2 \end{array} \right) (1-X^L_{\ell\ell'})$  \\ \hline
\end{tabular}}
\caption{
The expressions of $F^{L,p,X,zz}_{\ell\ell'}$ that appear in Eq.~(\ref{eq:orfzz}) for the ORFs for intensity (I), circular-polarization (V), and E- and B-mode linear-polarization anisotropies of multipole $L$.
See Eq.~(\ref{eq:z_t}) for $z^t_\ell$ and Eq.~(\ref{eq:z_vl}) for $z^v_\ell$. 
}
\label{tab:F_t_zz}
\end{table*}

\subsubsection{Spin-1 GWs}

\begin{figure}
        \centering \includegraphics[width=0.6\columnwidth]{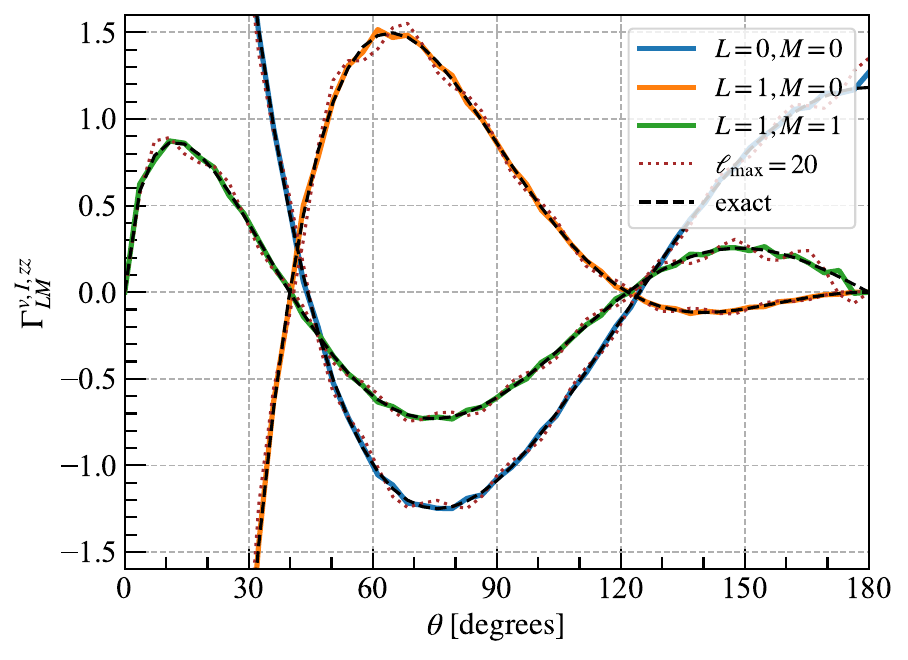}
        \caption{ The ORFs of $\Gamma^{v,I,zz}_{LM}(\theta)$.
        The color solid lines are for $\ell_\tmax = 40$.
        The brown dotted lines are for $\ell_\tmax = 20$. 
        The black dashed lines are the exact analytical results obtained in Ref.~\cite{Gair:2015hra}.
        Note that the $M=0$ cases diverge at $\theta =0$~\cite{Gair:2015hra}. 
    }
        \label{fig:g_v_I_zz}
\end{figure}

Next, we consider the spin-1 GWs. 
The procedure is basically the same as the case of spin-2 GWs.
From Eq.~(\ref{eq:z_nk}), the redshift response due to the spin-1 GWs with the frequency $f$ and the propagation direction $\hat k$ becomes
\begin{align}
  z(\hat n,\hat k) &= \frac{1}{2} \left[ \tilde h^x \frac{\sin 2 \theta \cos \phi}{1 + \cos\theta} + \tilde h^y \frac{\sin 2 \theta \sin \phi}{1 + \cos\theta}\right] \nonumber \\
  &= \frac{1}{2\sqrt{2}} \left[ \tilde h^v_L \frac{\sin 2 \theta }{1 + \cos\theta} \ee^{i\phi} + \tilde h^v_L\frac{\sin 2 \theta }{1 + \cos\theta} \ee^{- i \phi}\right].
  \label{eq:z_nk_s1}
\end{align}
From this, we can obtain 
\begin{align}
  \tilde z_{f,\ell m}(\hat k) &= \int \dd^2 \hat n \, Y^*_{\ell m}(\hat n) z_f(\hat n,\hat k) \nonumber \\
  & = \frac{z^v_\ell}{\sqrt{2}} \left( \tilde h^v_L(f,\hat k)\delta_{m1} - \tilde h^v_R(f,\hat k) \delta_{m,-1} \right),
  \label{eq:z_lm_v}
\end{align}
where 
\begin{align}
  z^v_\ell \equiv (-1)^{\ell+1} \sqrt{4\pi(2\ell + 1)} \left( \frac{1}{\sqrt{\ell(\ell+1)}} - \frac{\sqrt{2}}{3}\delta_{\ell1} \right).
  \label{eq:z_vl}
\end{align}
Similar to the spin-2 GW case, we can find $z_{f,\ell m} = 0$ in $\ell < 1$, which leads to $\ell_\tmin = 1$ in Eq.~(\ref{eq:orfzz}) for the spin-1 GWs.
We here note that the relative sign between the two terms in Eq.~(\ref{eq:z_lm_v}) is different from the spin-2 GW case (Eq.~(\ref{eq:z_lm_t})). Technically, this difference comes from the overall sign difference for $Y_{\ell 1}(\theta,0)$ and $Y_{\ell, -1}(\theta,0)$.
We here generalize this to the case of GWs propagating in $\hat \Omega$:
\begin{align}
  \tilde z_{f,\ell m}(\hat \Omega) &= \int \dd \hat n \, Y^*_{\ell m}(\hat n) z_f(\hat n,\hat \Omega) \nonumber \\
  & = \frac{z^v_\ell}{\sqrt{2}} \left(\tilde h^v_{L}(f,\hat \Omega) D^{(\ell')}_{m',1} - \tilde h^v_R(f,\hat \Omega) D^{(\ell')}_{m',-1} \right).
  \label{eq:z_lm_v2}
\end{align}
Using this, we finally obtain 
\begin{align}
  \vev{z^*_{f,\ell m} z_{f,\ell' m'} } &= \frac{z^v_\ell z^v_{\ell'}}{2} \int \dd^2 \hat\Omega \int \dd^2 \hat\Omega' \left[ \tilde h^{v\,,*}_L(f,\hat \Omega) D^{(\ell)*}_{m,1} - \tilde h^{v\,*}_R(f,\hat \Omega) D^{(\ell)*}_{m,-1} \right] \left[ \tilde h^v_{L}(f,\hat \Omega') D^{(\ell')}_{m',1} - \tilde h^v_R(f,\hat \Omega') D^{(\ell')}_{m',-1} \right] \nonumber \\
  &= \frac{z^v_\ell z^v_{\ell'}}{2} I^v_0(f) \sum_{LM} \left[ c^{v,I}_{LM}(f) \left( \,^{(0,1,1)}W^{LM}_{\ell \ell' m m'} + \,^{(0,-1,-1)}W^{LM}_{\ell \ell' m m'}\right) + c^{v,V}_{LM}(f) \left(\,^{(0,1,1)}W^{LM}_{\ell \ell' m m'} - \,^{(0,-1,-1)}W^{LM}_{\ell \ell' m m'} \right) \right. \nonumber \\
  & \quad \qquad\qquad\qquad \left.
  - c^{v,+}_{LM}(f) \,^{(2,1,-1)}W^{LM}_{\ell \ell' m m'} - c^{v,-}_{LM}(f) \,^{(-2,-1,1)}W^{LM}_{\ell \ell' m m'} \right] \nonumber \\  
  &=  2 z^v_\ell z^v_{\ell'} I^v_0(f) \sum_{LM}  (-1)^{m+L}\sqrt{\pi} \Braket{\ell (-m) \ell' m' |L M} \nonumber \\
  &\qquad
  \times\left\{ 
  \left[-c^{v,I}_{LM}(f) X^L_{\ell \ell'} + c^{v,V}_{LM}(f) (1-X^L_{\ell \ell'}) \right]
\begin{pmatrix}
  \ell & \ell' & L\\
  -1 & 1 & 0\\
  \end{pmatrix} 
  + 
  \left[c^{v,E}_{LM}(f) X^L_{\ell \ell'} + i c^{v,B}_{LM}(f) (1-X^L_{\ell \ell'}) \right]
\begin{pmatrix}
  \ell & \ell' & L\\
  1 & 1 & -2\\
  \end{pmatrix} 
  \right\}.
  \label{eq:z_lm_corr_v}
\end{align}
Substituting this into Eq.~(\ref{eq:zz_corr}), we obtain the coefficient $F^L_{\ell \ell'}$, summarized in the right column of Table~\ref{tab:F_t_zz}.
We here note that the sign difference from Eq.~(\ref{eq:z_lm_corr_t}) comes from the combination of the relative sign in Eq.~(\ref{eq:z_lm_v}) and the overall additional minus sign in Eq.~(\ref{eq:z_lm_corr_v}), which originates from the odd integer $s$ in $D^{(\ell)}_{m,s}$ in Eq.~(\ref{eq:z_lm_v2}).
The same sign difference between the spin-2 and the spin-1 GW cases appears in the following subsections (see Tables~\ref{tab:F_tb}-\ref{tab:F_bb}) and can be explained in the same way.
Figure~\ref{fig:g_v_I_zz} shows $\Gamma^{v,I,zz}_{LM}(\theta)$ with different $\ell_\tmax$ in the coordinates of Eq.~(\ref{eq:coord}).
In the figure, We also show the exact analytical results in Ref.~\cite{Gair:2014rwa}. 
The convergence with respect to $\ell_\tmax$ is slower than the spin-2 GW case (Fig.~\ref{fig:g_t_I_zz}). 
This is because $z^v_{\ell}$ decays slower than $z^t_{\ell}$ in $\ell \gg 1$ ($z^v_{\ell} \propto \ell^{-1/2}$ and $z^t_{\ell} \propto \ell^{-3/2}$ in $\ell \gg 1$).

\subsection{Auto-correlation of deflection fields}
\label{subsec:auto_def}

\begin{figure}
        \centering \includegraphics[width=0.5\columnwidth]{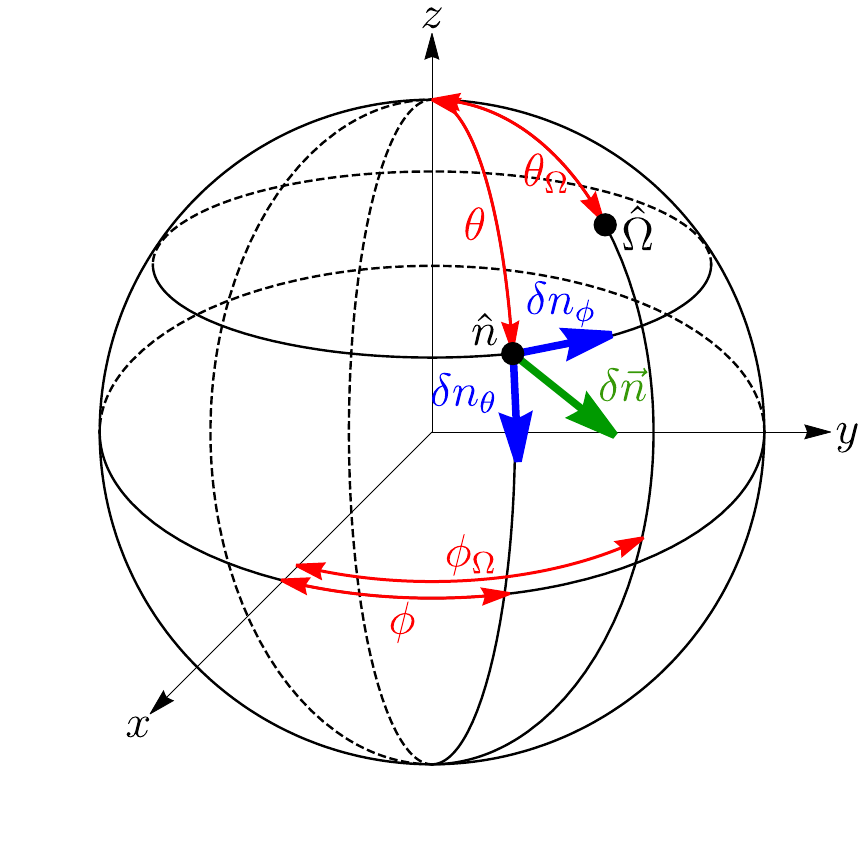}
        \caption{ The angles and the deflection vector field. 
    }
        \label{fig:sphere}
\end{figure}

Next, we consider the angular deflection due to GWs, expressed as $\hat {n}_\text{obs} = \hat n + \delta \vec n(\hat n)$ with $\hat {n}_\text{obs}$ being the observed celestial angle for stars or quasars and $\hat n$ being that in the case without GWs.
The angular deflection field caused by GWs propagating in $\hat \Omega$ is given by~\cite{Book:2010pf}
\begin{align}
  \delta n^i(t,\hat n,\hat\Omega) = \frac{(\hat n^i + \hat \Omega^i) \hat n^j \hat n^k h_{jk}(t,\hat \Omega,\vec 0)}{2(1 + \hat n \cdot \hat \Omega)} - \frac{1}{2} \hat n^j h_{ij}(t,\hat \Omega,\vec 0),
  \label{eq:n_deflec}
\end{align}
where the origin $\vec 0$ corresponds to the observer position.
In the following, we raise or lower the spatial indices with the Kronecker delta, e.g. $\delta n^i = \delta^{ij} \delta n_j$.
Figure~\ref{fig:sphere} visualizes the angles and the deflection vector field.
We expand the deflection vector $\delta \vec n$ as 
\begin{align}
    \delta n_i(t,\hat n,\hat\Omega) &= \sum_{\ell m} \left[ E_{\ell m}(t,\hat \Omega) Y^E_{(\ell m)i}(\hat n) + B_{\ell m}(t,\hat \Omega) Y^B_{(\ell m)i}(\hat n) \right],
\end{align}
where we take the same notation of the vector spherical harmonics as Refs.~\cite{Dai:2012bc,Qin:2018yhy}:
\begin{align}
  \label{eq:y_e_lm}
  Y^E_{(\ell m) i}(\theta, \phi) &= -\frac{r}{\sqrt{\ell(\ell + 1)}} \nabla_i Y_{\ell m}(\theta, \phi) = - \frac{1}{\sqrt{\ell(\ell + 1)}} \left[\hat \theta_i \frac{\partial}{\partial \theta} + \hat \phi_i \frac{1}{\sin \theta} \frac{\partial}{\partial \phi} \right] Y_{\ell m} (\theta, \phi) \nonumber \\
    &= - \frac{1}{2\sqrt{\ell (\ell + 1)}} \biggr\{ \hat \theta_i\left[ \sqrt{(\ell-m)(\ell+m+1)} \ee^{-i \phi} Y_{\ell, m+1}(\theta, \phi) - \sqrt{(\ell+m)(\ell-m+1)} \ee^{i \phi} Y_{\ell, m-1}(\theta, \phi) \right]\nonumber \\ 
  & \qquad \qquad \qquad \qquad 
  + \hat \phi_i \frac{2im}{\sin \theta} Y_{\ell m}(\theta, \phi) \biggl\}, \\
  \label{eq:y_b_lm} 
  Y^B_{(\ell m) i}(\theta, \phi) &= -\frac{r}{\sqrt{\ell(\ell + 1)}} \epsilon_{ijk} \hat n^j \nabla^k Y_{\ell m}(\theta, \phi) = - \frac{1}{\sqrt{\ell(\ell + 1)}} \left[\hat \phi_i \frac{\partial}{\partial \theta} - \hat \theta_i \frac{1}{\sin \theta} \frac{\partial}{\partial \phi} \right] Y_{\ell m} (\theta, \phi) \nonumber \\
  &= - \frac{1}{2\sqrt{\ell (\ell + 1)}} \biggr\{ \hat \phi_i\left[ \sqrt{(\ell-m)(\ell+m+1)} \ee^{-i \phi} Y_{\ell, m+1}(\theta, \phi) - \sqrt{(\ell+m)(\ell-m+1)} \ee^{i \phi} Y_{\ell, m-1}(\theta, \phi) \right]\nonumber \\ 
  & \qquad \qquad \qquad \qquad 
  - \hat \theta_i \frac{2im}{\sin \theta} Y_{\ell m}(\theta, \phi) \biggl\},
\end{align}
where $r$ is the radial distance, which is cancelled by $r^{-1}$ in $\nabla_i$, and $\hat \theta$ and $\hat \phi$ are the unit vectors satisfying $\hat \theta \cdot \hat n = \hat \phi \cdot \hat n = 0$ and being along the $\theta$ and the $\phi$ variation, respectively.
Note the above vector spherical harmonics live on the surface perpendicular to $\hat n$.
We define the spatial indices as those satisfying $\hat \theta_\theta = \hat \phi_\phi = 1$ and $\hat \theta_\phi = \hat \phi_\theta = 0$.
With the additional spatial derivative, we find the following relation satisfied:
\begin{align}
  r \nabla^i Y^E_{(\ell m) i} &= -\frac{r^2}{\sqrt{\ell(\ell + 1)}} \nabla^2 Y_{\ell m}(\theta, \phi) = \sqrt{\ell(\ell + 1)} Y_{\ell m}(\theta, \phi), \\
  r(\epsilon_{ijk} \hat n^j \nabla^k) Y^B_{(\ell m) i} &= -\frac{r^2}{\sqrt{\ell(\ell + 1)}} \nabla^2 Y_{\ell m}(\theta, \phi) = \sqrt{\ell(\ell + 1)} Y_{\ell m}(\theta, \phi),
\end{align}
where we have used $r^2 \nabla^2 Y_{\ell m} = -\ell (\ell + 1) Y_{\ell m}$.
Using this, we define 
\begin{align}
  \bar E(\hat n) &\equiv r \nabla_i \delta n^i = \sum_{\ell m} \bar E_{\ell m} Y_{\ell m}, \\
  \bar B(\hat n) &\equiv r(\epsilon_{ijk} \hat n^j \nabla^k) \delta n^i = \sum_{\ell m} \bar B_{\ell m} Y_{\ell m}, 
\end{align}
where 
\begin{align}
\label{eq:bar_non_bar_eb}
\bar E_{\ell m} \equiv \sqrt{\ell(\ell + 1)} E_{\ell m}, \ \bar B_{\ell m} \equiv \sqrt{\ell(\ell + 1)} B_{\ell m}.
\end{align}
From Eq.~(\ref{eq:n_deflec}), we can also define a scalar $\bar E(t,\hat n)$ and pseudoscalar $\bar B(t,\hat n)$ induced by GWs propagating in $\hat \Omega$ as 
\begin{align}
  \label{eq:til_e}
  \bar E(t,\hat n,\hat\Omega)&= - \frac{1}{2} \text{Tr}\, h(t,\hat \Omega,\vec 0) + \frac{(\hat n^i + \hat \Omega^i) \hat n^j h_{ij}(t,\hat \Omega,\vec 0)}{1 + \hat \Omega \cdot \hat n}, \\ 
  \label{eq:til_b} 
  \bar B(t,\hat n,\hat\Omega) &= \epsilon^{ijk} \frac{\hat \Omega_i \hat n^l \hat n_k h_{jl}(t,\hat \Omega,\vec 0)}{1 + \hat \Omega \cdot \hat n},
\end{align}
where we have used $r\nabla_i \hat n_j = \delta_{ij} - \hat n_i \hat n_j$.

Similar to the redshift response case (Eq.~(\ref{eq:zz_corr})), we consider the two-point correlation function:
\begin{align}
  \vev{\delta n^*_{f,i}(\hat n_a) \delta n_{f,j}(\hat n_b)} &= \sum_{\ell m} \sum_{\ell' m'} \sum_{S,T}\vev{S^*_{f,\ell m} T_{f,\ell' m'}} Y^{S*}_{(\ell m),i}(\hat n_a) Y^T_{(\ell' m'),j}(\hat n_b) \nonumber \\
  &= \sum_{p=t,v} \sum_X \sum_{S,T} i^X \left[ I^p_0(f) \sum_{L,M} c^{p,X}_{LM}(f) \Gamma^{p,X,ST}_{LM,ij}(\hat n_a, \hat n_b) \right],
  \label{eq:nn_corr}
\end{align}
where $S,T \in \{E,B\}$. 
We note that there are two types of $E$ and $B$ modes here: those for the polarization of GWs and those for the deflection angle. 
Then, the ORF is given by 
\begin{tcolorbox}[ams align]
  \Gamma^{p,X,ST}_{(LM)ij}(\hat n_a, \hat n_b) = (-1)^L \sqrt{\pi} \sum^{\ell_\tmax}_{\ell=\ell_\tmin} \sum^{\ell_\tmax}_{\ell' = \ell_\tmin} F^{L,p,X,ST}_{\ell \ell'}\left\{ Y^S_{(\ell)i}(\hat n_a) \otimes Y^T_{(\ell')j}(\hat n_b)\right\}_{LM},
  \label{eq:orf}
\end{tcolorbox}  
where 
\begin{align}
  \left\{ Y^S_{(\ell)i}(\hat n_a) \otimes Y^T_{(\ell')j}(\hat n_b)\right\}_{LM} \equiv \sum^{\ell}_{m=-\ell} \sum^{\ell'}_{m'=-\ell'} \Braket{\ell m \ell' m' |L M} Y^S_{(\ell m)i}(\hat n_a)Y^T_{(\ell' m')j}(\hat n_b).
\end{align}
We will obtain the expression of $F^{L,p,X,ST}_{\ell \ell'}$ in the following.

\subsubsection{Spin-2 GWs}

We begin with spin-2 GWs. 
The main goal of this subsection is to calculate $\braket{S^*_{f,\ell m} T_{f,\ell' m'}}$ and obtain their ORF. 
For convenience, we first obtain $\bar E_{\ell m}$ and $\bar B_{\ell m}$ and then convert them to $E_{\ell m}$ and $B_{\ell m}$.
From Eqs.~(\ref{eq:til_e}) and (\ref{eq:til_b}), the $E$- and $B$-modes of the deflection due to the spin-2 GWs with the frequency $f$ and the propagation direction $\hat k$ become
\begin{align}
  \label{eq:bar_e_f}
  \bar E_f(\hat n,\hat k) &= \tilde h_+(1- \cos\theta) \cos(2 \phi) + \tilde h_\times (1- \cos\theta) \sin(2 \phi) \nonumber \\
  &= \frac{1}{\sqrt{2}} \left[ \tilde h^t_L (1- \cos\theta) \ee^{2i \phi} + \tilde h^t_R (1- \cos\theta) \ee^{-2i\phi}\right], \\
  \bar B_f(\hat n,\hat k) &= \tilde h_+(1-\cos \theta) \sin 2\phi \,  - \tilde h_\times(1-\cos \theta) \cos 2\phi \nonumber \\
  &= \frac{-i}{\sqrt{2}} \left[ \tilde h^t_L (1- \cos\theta) \ee^{2i \phi} - \tilde h^t_R (1- \cos\theta) \ee^{-2i\phi}\right],
\end{align}
where we have taken again the coordinates of $\hat k = (0,0,1)$ and $\hat n = (\sin \theta \cos \phi, \sin\theta \sin \phi, \cos \theta)$.
We note that the right-hand side of Eq.~(\ref{eq:bar_e_f}) is the same as Eq.~(\ref{eq:z_nk2}) except for the overall factor $2$.
From this and Eq.~(\ref{eq:bar_non_bar_eb}), we can obtain 
\begin{align}
  \tilde E_{f,\ell m}(\hat k) &= \frac{1}{\sqrt{\ell(\ell+1)}}\int \dd^2 \hat n \, Y^*_{\ell m}(\hat n) \bar E_f(\hat n,\hat k) \nonumber \\
  & = \frac{E^t_\ell}{\sqrt{2}} \left( \tilde h^t_L(f,\hat k)\delta_{m2} + \tilde h^t_R(f,\hat k) \delta_{m,-2} \right),
  \label{eq:e_lm_t}\\
  \tilde B_{f,\ell m}(\hat k) &= \frac{1}{\sqrt{\ell(\ell+1)}}\int \dd^2 \hat n \, Y^*_{\ell m}(\hat n) \bar B_f(\hat n,\hat k) \nonumber \\
  & = \frac{B^t_\ell}{\sqrt{2}} \left( \tilde h^t_L(f,\hat k) \delta_{m2} - \tilde h^t_R(f, \hat k)\delta_{m,-2} \right),
  \label{eq:e_lm_t}
\end{align}
where 
\begin{align}
  E^t_\ell \equiv \frac{2 (-1)^\ell}{\sqrt{\ell(\ell+1)}} \sqrt{\frac{4\pi (2\ell + 1)(\ell-2)!}{(\ell+2)!}} = \frac{2}{\sqrt{\ell(\ell+1)}} z^t_\ell, \ \ 
  B^t_\ell \equiv (-i) E^t_\ell.
  \label{eq:eb_t_l}
\end{align}
Similar to the redshift response case, we can obtain the expression for GWs propagating in a general direction $\hat \Omega$:
\begin{align}
  \tilde E_{f,\ell m}(\hat \Omega) & = \frac{E^t_\ell}{\sqrt{2}} \left(\tilde h^t_{L}(f,\hat \Omega) D^{(\ell)}_{m,2} + \tilde h^t_R(f,\hat \Omega) D^{(\ell)}_{m,-2} \right),
  \label{eq:e_lm_t2}\\
  \tilde B_{f,\ell m}(\hat \Omega) & = \frac{B^t_\ell}{\sqrt{2}} \left(\tilde h^t_{L}(f,\hat \Omega) D^{(\ell)}_{m,2} - \tilde h^t_R(f,\hat \Omega) D^{(\ell)}_{m,-2} \right).  
\end{align}
We here define $S_{f,\ell m} = \int \dd^2 \hat \Omega\, \tilde S_{f,\ell m}(\hat \Omega)$ ($S \in \{E,B\}$) and obtain 
\begin{align}
  \vev{E^*_{f,\ell m} E_{f,\ell' m'} } &= \frac{E^t_\ell E^t_{\ell'}}{z^t_\ell z^t_{\ell'}} \vev{z^*_{f,\ell m} z_{f,\ell' m'}} = \frac{4}{\sqrt{\ell (\ell + 1)\ell'(\ell'+1)}} \vev{z^*_{f,\ell m} z_{f,\ell' m'}}, \\
  \vev{E^*_{f,\ell m} B_{f,\ell' m'} } &= \frac{E^t_\ell B^t_{\ell'}}{2} \int \dd^2 \hat\Omega \int \dd^2 \hat\Omega' \left[ \tilde h^{t\,,*}_L(f,\hat \Omega) D^{(\ell)*}_{m,2} + \tilde h^{t\,*}_R(f,\hat \Omega) D^{(\ell)*}_{m,-2} \right] \left[ \tilde h^t_{L}(f,\hat \Omega') D^{(\ell')}_{m',2} - \tilde h^t_R(f,\hat \Omega') D^{(\ell')}_{m',-2} \right] \nonumber \\
  &=  2 E^t_\ell B^t_{\ell'} I^t_0(f) \sum_{LM}  (-1)^{m+L}\sqrt{\pi} \Braket{\ell (-m) \ell' m' |L M} \nonumber \\
  &\qquad
  \times\left\{ 
  \left[-c^{t,I}_{LM}(f) (1-X^L_{\ell \ell'}) + c^{t,V}_{LM}(f) X^L_{\ell \ell'} \right]
\begin{pmatrix}
  \ell & \ell' & L\\
  -2 & 2 & 0\\
  \end{pmatrix} 
  - 
  \left[c^{t,E}_{LM}(f) (1-X^L_{\ell \ell'}) + i c^{t,B}_{LM}(f) X^L_{\ell \ell'} \right]
\begin{pmatrix}
  \ell & \ell' & L\\
  2 & 2 & -4\\
  \end{pmatrix} 
  \right\},\\
  \vev{B^*_{f,\ell m} E_{f,\ell' m'} } &= \frac{B^{t*}_\ell E^t_{\ell'}}{2} \int \dd^2 \hat\Omega \int \dd^2 \hat\Omega' \left[ \tilde h^{t\,,*}_L(f,\hat \Omega) D^{(\ell)*}_{m,2} - \tilde h^{t\,*}_R(f,\hat \Omega) D^{(\ell)*}_{m,-2} \right] \left[ \tilde h^t_{L}(f,\hat \Omega') D^{(\ell')}_{m',2} + \tilde h^t_R(f,\hat \Omega') D^{(\ell')}_{m',-2} \right] \nonumber \\
  &=  2 B^{t*}_\ell E^t_{\ell'} I^t_0(f) \sum_{LM}  (-1)^{m+L}\sqrt{\pi} \Braket{\ell (-m) \ell' m' |L M} \nonumber \\
  &\qquad
  \times\left\{ 
  \left[-c^{t,I}_{LM}(f) (1-X^L_{\ell \ell'}) + c^{t,V}_{LM}(f) X^L_{\ell \ell'} \right]
\begin{pmatrix}
  \ell & \ell' & L\\
  -2 & 2 & 0\\
  \end{pmatrix} 
  + 
  \left[c^{t,E}_{LM}(f) (1-X^L_{\ell \ell'}) + i c^{t,B}_{LM}(f) X^L_{\ell \ell'} \right]
\begin{pmatrix}
  \ell & \ell' & L\\
  2 & 2 & -4\\
  \end{pmatrix} 
  \right\}, \\
  \vev{B^*_{f,\ell m} B_{f,\ell' m'} } &= \frac{B^{t*}_\ell B^t_{\ell'}}{2} \int \dd^2 \hat\Omega \int \dd^2 \hat\Omega' \left[ \tilde h^{t\,,*}_L(f,\hat \Omega) D^{(\ell)*}_{m,2} - \tilde h^{t\,*}_R(f,\hat \Omega) D^{(\ell)*}_{m,-2} \right] \left[ \tilde h^t_{L}(f,\hat \Omega') D^{(\ell')}_{m',2} - \tilde h^t_R(f,\hat \Omega') D^{(\ell')}_{m',-2} \right] \nonumber \\
  &=  2 B^{t*}_\ell B^t_{\ell'} I^t_0(f) \sum_{LM}  (-1)^{m+L}\sqrt{\pi} \Braket{\ell (-m) \ell' m' |L M} \nonumber \\
  &\qquad
  \times\left\{ 
  \left[c^{t,I}_{LM}(f) X^L_{\ell \ell'} - c^{t,V}_{LM}(f) (1-X^L_{\ell \ell'}) \right]
\begin{pmatrix}
  \ell & \ell' & L\\
  -2 & 2 & 0\\
  \end{pmatrix} 
  - 
  \left[c^{t,E}_{LM}(f) X^L_{\ell \ell'} + i c^{t,B}_{LM}(f) (1-X^L_{\ell \ell'}) \right]
\begin{pmatrix}
  \ell & \ell' & L\\
  2 & 2 & -4\\
  \end{pmatrix} 
  \right\},
\end{align}
where see Eq.~(\ref{eq:z_lm_corr_t}) for the expression of $\braket{z^*_{f,\ell m} z_{f,\ell' m'}}$ due to spin-2 GWs.

\begin{figure}
        \centering \includegraphics[width=0.9\columnwidth]{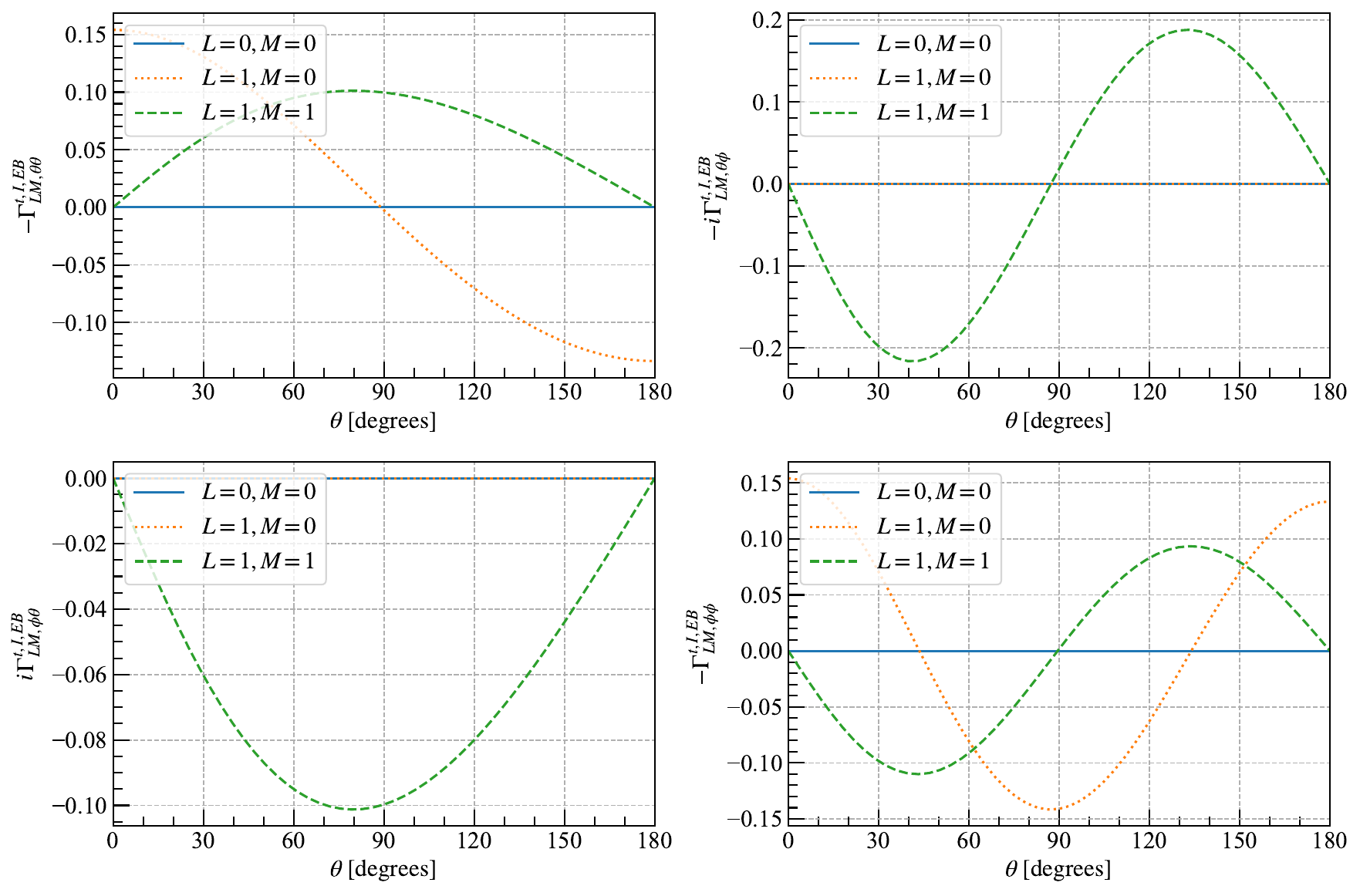}
        \caption{ The ORFs of $\Gamma^{t,I,EB}_{LM}(\theta)$ with $\ell_\tmax = 10$.
    }
        \label{fig:g_t_I_eb}
\end{figure}

From these, the ORF coefficients are
\begin{align}
  F^{L,t,X,EE}_{\ell \ell'} = \frac{4}{\sqrt{\ell(\ell + 1) \ell' (\ell' + 1)}} F^{L,t,X,zz}_{\ell \ell'},
\end{align}
and the left column in Tables~\ref{tab:F_tb} for $F^{L,t,X,EB}_{\ell \ell'}$, \ref{tab:F_bt} for $F^{L,t,X,BE}_{\ell \ell'}$, and \ref{tab:F_bb} for $F^{L,t,X,BB}_{\ell \ell'}$.
To get the concrete plots of the ORFs, we similarly take the coordinates given in Eq.~(\ref{eq:coord}). 
Strictly speaking, in that choice of coordinates, $Y^E_{(\ell m) \phi}(\hat n_a)$ and $Y^B_{(\ell m)\theta}(\hat n_a)$ are not well defined because of $\sin \theta$ in the denominator, which is from the singularity of the coordinates.
To avoid this issue, we define $Y^E_{(\ell m) \phi}(\hat n_a)$ and $Y^B_{(\ell m)\theta}(\hat n_a)$ with $\hat n_a = \hat n_b|_{\theta \rightarrow +0}$ in the coordinates. Practically, we take $\hat n_a = (\sin \theta_\epsilon, 0, \cos \theta_\epsilon)$ with $\theta_\epsilon = 10^{-5}$ in the numerical calculation for Figs.~\ref{fig:g_t_I_eb}-\ref{fig:g_v_I_ez}.
Then, Figure~\ref{fig:g_t_I_eb} shows the $\theta$-dependence of $\Gamma^{t,I,EB}_{LM}$.
In the figure, we can see $\Gamma^{t,I,EB}_{00} =0$. This physically means that the parity-breaking signal ($EB$-correlation) cannot be produced in the parity-conserving background (isotropic and unpolarized background).

\begin{table*}
\centering
    \everymath{\displaystyle}
    \resizebox{0.8\textwidth}{!}{\begin{tabular}{|c|c|c|}
    \hline
    $X$ & Spin-2 & Spin-1  \\
  \hline\hline

    $I$ & $-2 E_\ell^{t} B_{\ell'}^t  \left( \begin{array}{ccc} \ell & \ell'& L \\ -2 & 2 &  0 \end{array} \right) (1-X^L_{\ell\ell'})$ &  $2 E_\ell^{v} B_{\ell'}^v  \left( \begin{array}{ccc} \ell & \ell'& L \\ -1 & 1 &  0 \end{array} \right) (1-X^L_{\ell\ell'})$  \\ \hline

    $V$ & $2 E_\ell^{t} B_{\ell'}^t  \left( \begin{array}{ccc} \ell & \ell'& L \\ -2 & 2 &  0 \end{array} \right) X^L_{\ell\ell'}$ &  $-2E_\ell^{v} B_{\ell'}^v  \left( \begin{array}{ccc} \ell & \ell'& L \\ -1 & 1 &  0 \end{array} \right) X^L_{\ell\ell'}$  \\ \hline
    
    $E$ & $-2 E_\ell^{t} B_{\ell'}^t  \left( \begin{array}{ccc} \ell & \ell'& L \\ 2 & 2 &  -4 \end{array} \right) (1-X^L_{\ell\ell'})$ &  $-2 E_\ell^{v} B_{\ell'}^v  \left( \begin{array}{ccc} \ell & \ell'& L \\ 1 & 1 &  -2 \end{array} \right) (1-X^L_{\ell\ell'})$  \\ \hline

    $B$ & $-2 E_\ell^{t} B_{\ell'}^t  \left( \begin{array}{ccc} \ell & \ell'& L \\ 2 & 2 &  -4 \end{array} \right) X^L_{\ell\ell'}$  &  $ -2 E_\ell^{v} B_{\ell'}^v  \left( \begin{array}{ccc} \ell & \ell'& L \\ 1 & 1 &  -2 \end{array} \right) X^L_{\ell\ell'}$  \\ \hline
\end{tabular}}
\caption{
The summary of $F^{L,p,X,EB}_{\ell \ell'}$.
See Eq.~(\ref{eq:eb_t_l}) for $E^t_\ell$ and $B^t_\ell$ and Eq.~(\ref{eq:eb_vl}) for $E^v_\ell$ and $B^v_\ell$.
}
\label{tab:F_tb}
\end{table*}

\begin{table*}
\centering
    \everymath{\displaystyle}
    \resizebox{0.8\textwidth}{!}{\begin{tabular}{|c|c|c|}
    \hline
    $X$ & Spin-2 & Spin-1  \\
  \hline\hline

    $I$ & $-2 B_\ell^{t\,*} E_{\ell'}^t  \left( \begin{array}{ccc} \ell & \ell'& L \\ -2 & 2 &  0 \end{array} \right) (1-X^L_{\ell\ell'})$ &  $2 B_\ell^{v\,*} E_{\ell'}^v  \left( \begin{array}{ccc} \ell & \ell'& L \\ -1 & 1 &  0 \end{array} \right) (1-X^L_{\ell\ell'})$  \\ \hline

    $V$ & $2 B_\ell^{t\,*} E_{\ell'}^t  \left( \begin{array}{ccc} \ell & \ell'& L \\ -2 & 2 &  0 \end{array} \right) X^L_{\ell\ell'}$ &  $-2B_\ell^{v\,*} E_{\ell'}^v  \left( \begin{array}{ccc} \ell & \ell'& L \\ -1 & 1 &  0 \end{array} \right) X^L_{\ell\ell'}$  \\ \hline
    
    $E$ & $2 B_\ell^{t\,*} E_{\ell'}^t  \left( \begin{array}{ccc} \ell & \ell'& L \\ 2 & 2 &  -4 \end{array} \right) (1-X^L_{\ell\ell'})$ &  $2 B_\ell^{v\,*} E_{\ell'}^v  \left( \begin{array}{ccc} \ell & \ell'& L \\ 1 & 1 &  -2 \end{array} \right) (1-X^L_{\ell\ell'})$  \\ \hline

    $B$ & $2 B_\ell^{t\,*} E_{\ell'}^t  \left( \begin{array}{ccc} \ell & \ell'& L \\ 2 & 2 &  -4 \end{array} \right) X^L_{\ell\ell'}$  &  $ 2 B_\ell^{v\,*} E_{\ell'}^v  \left( \begin{array}{ccc} \ell & \ell'& L \\ 1 & 1 &  -2 \end{array} \right) X^L_{\ell\ell'}$  \\ \hline
\end{tabular}}
\caption{
The summary of $F^{L,p,X,BE}_{\ell \ell'}$.
}
\label{tab:F_bt}
\end{table*}

\begin{table*}
\centering
    \everymath{\displaystyle}
    \resizebox{0.8\textwidth}{!}{\begin{tabular}{|c|c|c|}
    \hline
    $X$ & Spin-2 & Spin-1  \\
  \hline\hline

    $I$ & $2 B_\ell^{t\,*} B_{\ell'}^t  \left( \begin{array}{ccc} \ell & \ell'& L \\ -2 & 2 &  0 \end{array} \right) X^L_{\ell\ell'}$ &  $-2 B_\ell^{v\,*} B_{\ell'}^v \left( \begin{array}{ccc} \ell & \ell'& L \\ -1 & 1 &  0 \end{array} \right) X^L_{\ell\ell'}$  \\ \hline
 
    $V$ & $ -2 B_\ell^{t\,*} B_{\ell'}^t  \left( \begin{array}{ccc} \ell & \ell'& L \\ -2 & 2 &  0 \end{array} \right) (1-X^L_{\ell\ell'})$ &  $2 B_\ell^{v\,*} B_{\ell'}^v \left( \begin{array}{ccc} \ell & \ell'& L \\ -1 & 1 &  0 \end{array} \right) (1-X^L_{\ell\ell'})$  \\ \hline
    
    $E$ & $-2 B_\ell^{t\,*} B_{\ell'}^t  \left( \begin{array}{ccc} \ell & \ell'& L \\ 2 & 2 &  -4 \end{array} \right) X^L_{\ell\ell'}$ &  $-2 B_\ell^{v\,*} B_{\ell'}^v \left( \begin{array}{ccc} \ell & \ell'& L \\ 1 & 1 &  -2 \end{array} \right) X^L_{\ell\ell'}$  \\ \hline

    $B$ & $-2 B_\ell^{t\,*} B_{\ell'}^t  \left( \begin{array}{ccc} \ell & \ell'& L \\ 2 & 2 &  -4 \end{array} \right) (1-X^L_{\ell\ell'})$  &  $ -2 B_\ell^{v\,*} B_{\ell'}^v \left( \begin{array}{ccc} \ell & \ell'& L \\ 1 & 1 &  -2 \end{array} \right) (1-X^L_{\ell\ell'})$  \\ \hline
\end{tabular}}
\caption{
The summary of $F^{L,p,X,BB}_{\ell \ell'}$.
}
\label{tab:F_bb}
\end{table*}

\subsubsection{Spin-1 GWs}

\begin{figure}
        \centering \includegraphics[width=0.9\columnwidth]{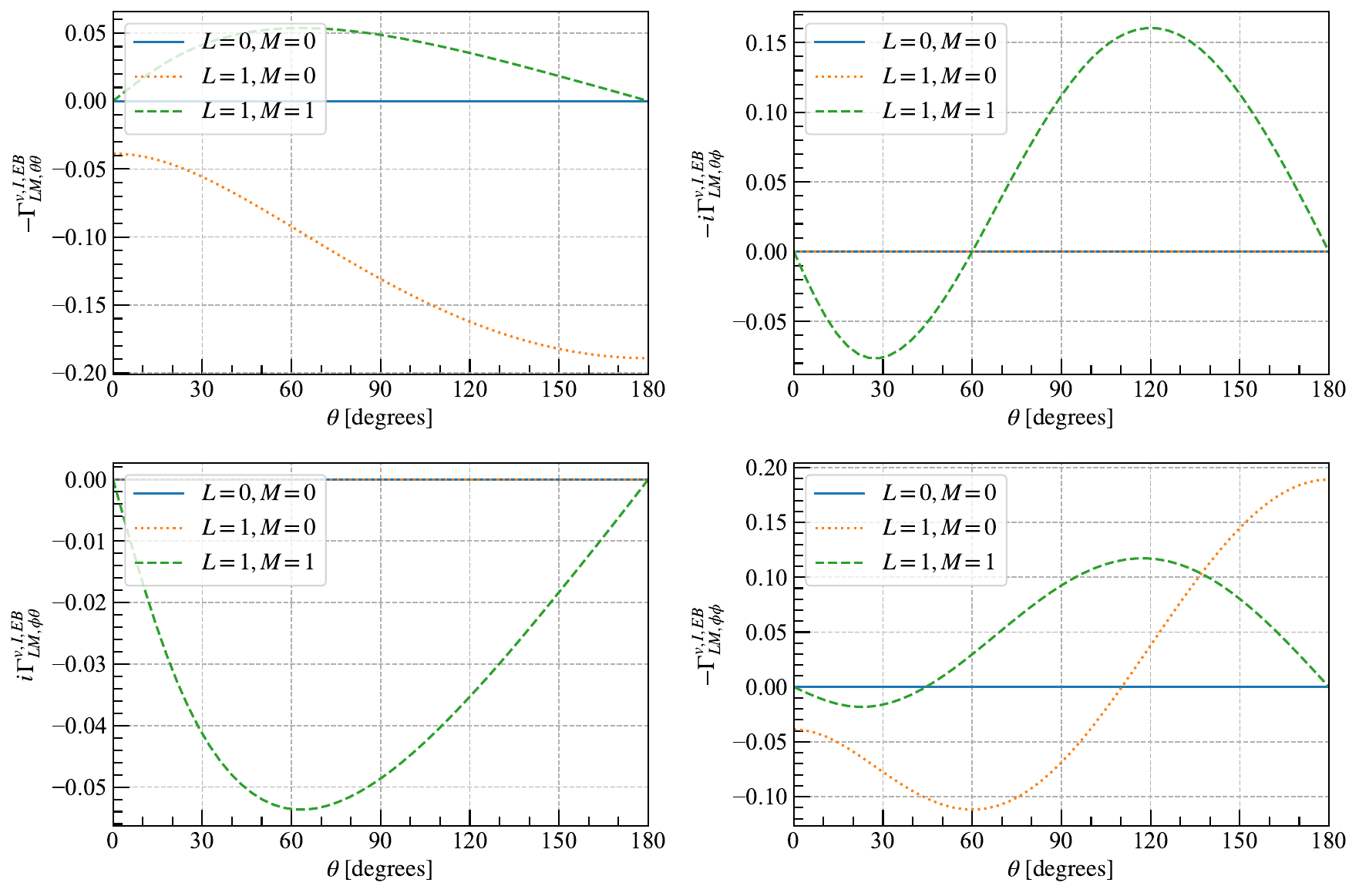}
        \caption{The ORFs of $\Gamma^{v,I,EB}_{LM}(\theta)$ with $\ell_\tmax = 10$.
    }
        \label{fig:g_v_I_eb}
\end{figure}

Next, we discuss the deflection due to spin-1 GWs.
From Eqs.~(\ref{eq:til_e}) and (\ref{eq:til_b}), the deflection due to the spin-1 GWs with the frequency $f$ and the propagation direction $\hat k$ becomes
\begin{align}
  \label{eq:e_spin1_ast}
  \bar E(\hat n,\hat k) &= \tilde h_x \frac{(\sin 2\theta + \sin \theta) \cos \phi}{1+\cos \theta} + \tilde h_y \frac{(\sin 2\theta + \sin \theta) \sin \phi}{1+\cos \theta} \nonumber \\
  &= \frac{1}{\sqrt{2}} \left[ \tilde h^v_L \frac{(\sin 2\theta + \sin \theta)}{1+\cos \theta} \ee^{i \phi} + \tilde h^v_R \frac{(\sin 2\theta + \sin \theta)}{1+\cos \theta} \ee^{-i \phi} \right], \\
  \bar B(\hat n,\hat k) &= \tilde h_x \frac{\sin 2\theta \sin \phi}{2(1+\cos \theta)} - \tilde h_y \frac{\sin 2\theta \cos \phi}{2(1+\cos \theta)} \nonumber \\
  &= \frac{1}{2\sqrt{2}} (-i) \left[ \tilde h^v_L \frac{\sin 2\theta}{1+\cos \theta} \ee^{i \phi} - \tilde h^v_R \frac{\sin 2\theta}{1+\cos \theta} \ee^{-i \phi} \right].
  \label{eq:b_spin1_ast}
\end{align}
We here note that Eq.~(\ref{eq:b_spin1_ast}) is the same as Eq.~(\ref{eq:z_nk_s1}) except for the relative sign between the two terms and the overall $(-i)$.
Then, we obtain 
\begin{align}
  \tilde E_{f,\ell m}(\hat k) &= \frac{1}{\sqrt{\ell(\ell+1)}} \int^{2\pi}_0 \dd \phi \int^\theta_0 \dd \theta \sin \theta\, Y^*_{\ell m}(\theta, \phi) \bar E(\hat n,\hat k) \nonumber \\
  & = \frac{E^v_\ell}{\sqrt{2}} \left( \tilde h^v_L(f,\hat k) \delta_{m1} - \tilde h^v_R(f,\hat k) \delta_{m,-1} \right),
  \label{eq:e_lm_s1}\\
  \tilde B_{f,\ell m}(\hat k) &= \frac{1}{\sqrt{\ell(\ell+1)}}\int^{2\pi}_0 \dd \phi \int^\theta_0 \dd \theta \sin \theta\, Y^*_{\ell m}(\theta, \phi) \bar B(\hat n,\hat k) \nonumber \\
  & = \frac{B^v_\ell}{\sqrt{2}} \left( \tilde h^v_L(f,\hat k) \delta_{m1} + \tilde h^v_R(f,\hat k) \delta_{m,-1} \right),
  \label{eq:b_lm_s1}
\end{align}
where 
\begin{align}
  E^v_\ell \equiv (-1)^{\ell+1} \sqrt{\frac{4\pi(2\ell + 1)}{\ell(\ell+1)}} \left( \frac{1}{\sqrt{\ell(\ell+1)}} - \frac{2\sqrt{2}}{3}\delta_{\ell1} \right),
  \ \ 
  B^v_\ell \equiv \frac{-i}{\sqrt{\ell(\ell+1)}} z^v_\ell.
  \label{eq:eb_vl}
\end{align}
We can calculate the ORF coefficients in the same way as the spin-1 case for the redshift response.
Similar to the previous cases, we can obtain the expression for GWs propagating in a general direction $\hat \Omega$:
\begin{align}
  \tilde E_{f,\ell m}(\hat \Omega) & = \frac{E^v_\ell}{\sqrt{2}} \left(\tilde h^v_{L}(f,\hat \Omega) D^{(\ell')}_{m',1} - \tilde h^v_R(f,\hat \Omega) D^{(\ell')}_{m',-1} \right),
  \label{eq:e_lm_v2}\\
  \tilde B_{f,\ell m}(\hat \Omega) & = \frac{B^v_\ell}{\sqrt{2}} \left(\tilde h^v_{L}(f,\hat \Omega) D^{(\ell')}_{m',1} + \tilde h^v_R(f,\hat \Omega) D^{(\ell')}_{m',-1} \right).
\end{align}
Note that Eq.~(\ref{eq:e_spin1_ast}) has the same structure as Eq.~(\ref{eq:z_lm_v2}) with $z^v_\ell \to E^v_\ell$.
We here define $S_{f,\ell m} = \int \dd^2 \hat \Omega\, \tilde S_{f,\ell m}(\hat \Omega)$ ($S \in \{E,B\}$) and obtain 
\begin{align}
  \vev{E^*_{f,\ell m} E_{f,\ell' m'} } &= \frac{E^v_\ell E^v_{\ell'}}{z^v_\ell z^v_{\ell'}} \vev{z^*_{f,\ell m} z_{f,\ell' m'}}, \\
  \vev{E^*_{f,\ell m} B_{f,\ell' m'} } &= \frac{E^v_\ell B^v_{\ell'}}{2} \int \dd^2 \hat\Omega \int \dd^2 \hat\Omega' \left[ \tilde h^{v\,,*}_L(f,\hat \Omega) D^{(\ell)*}_{m,1} - \tilde h^{v\,*}_R(f,\hat \Omega) D^{(\ell)*}_{m,-1} \right] \left[ \tilde h^v_{L}(f,\hat \Omega') D^{(\ell')}_{m',1} + \tilde h^v_R(f,\hat \Omega') D^{(\ell')}_{m',-1} \right] \nonumber \\
  &=  2 E^v_\ell B^v_{\ell'} I^v_0(f) \sum_{LM}  (-1)^{m+L}\sqrt{\pi} \Braket{\ell (-m) \ell' m' |L M} \nonumber \\
&\qquad
  \times\left\{ 
  \left[c^{v,I}_{LM}(f) (1-X^L_{\ell \ell'}) - c^{v,V}_{LM}(f) X^L_{\ell \ell'} \right]
\begin{pmatrix}
  \ell & \ell' & L\\
  -1 & 1 & 0\\
  \end{pmatrix} 
  + 
  \left[-c^{v,E}_{LM}(f) (1-X^L_{\ell \ell'}) - i c^{v,B}_{LM}(f) X^L_{\ell \ell'} \right]
\begin{pmatrix}
  \ell & \ell' & L\\
  1 & 1 & -2\\
  \end{pmatrix} 
  \right\},\\
  \vev{B^*_{f,\ell m} E_{f,\ell' m'} } &= \frac{B^{v*}_\ell E^v_{\ell'}}{2} \int \dd^2 \hat\Omega \int \dd^2 \hat\Omega' \left[ \tilde h^{v\,,*}_L(f,\hat \Omega) D^{(\ell)*}_{m,1} + \tilde h^{v\,*}_R(f,\hat \Omega) D^{(\ell)*}_{m,-1} \right] \left[ \tilde h^v_{L}(f,\hat \Omega') D^{(\ell')}_{m',1} - \tilde h^v_R(f,\hat \Omega') D^{(\ell')}_{m',-1} \right] \nonumber \\
  &=  2 B^{v*}_\ell E^v_{\ell'} I^v_0(f) \sum_{LM}  (-1)^{m+L}\sqrt{\pi} \Braket{\ell (-m) \ell' m' |L M} \nonumber \\
&\qquad
  \times\left\{ 
  \left[c^{v,I}_{LM}(f) (1-X^L_{\ell \ell'}) - c^{v,V}_{LM}(f) X^L_{\ell \ell'} \right]
\begin{pmatrix}
  \ell & \ell' & L\\
  -1 & 1 & 0\\
  \end{pmatrix} 
  + 
  \left[c^{v,E}_{LM}(f) (1-X^L_{\ell \ell'}) + i c^{v,B}_{LM}(f) X^L_{\ell \ell'} \right]
\begin{pmatrix}
  \ell & \ell' & L\\
  1 & 1 & -2\\
  \end{pmatrix} 
  \right\},\\
  \vev{B^*_{f,\ell m} B_{f,\ell' m'} } &= \frac{B^{v*}_\ell B^v_{\ell'}}{2} \int \dd^2 \hat\Omega \int \dd^2 \hat\Omega' \left[ \tilde h^{v\,,*}_L(f,\hat \Omega) D^{(\ell)*}_{m,1} + \tilde h^{v\,*}_R(f,\hat \Omega) D^{(\ell)*}_{m,-1} \right] \left[ \tilde h^v_{L}(f,\hat \Omega') D^{(\ell')}_{m',1} + \tilde h^v_R(f,\hat \Omega') D^{(\ell')}_{m',-1} \right] \nonumber \\
  &=  2 B^{v*}_\ell B^v_{\ell'} I^v_0(f) \sum_{LM}  (-1)^{m+L}\sqrt{\pi} \Braket{\ell (-m) \ell' m' |L M} \nonumber \\
&\qquad
  \times\left\{ 
  \left[-c^{v,I}_{LM}(f) X^L_{\ell \ell'} + c^{v,V}_{LM}(f) (1-X^L_{\ell \ell'}) \right]
\begin{pmatrix}
  \ell & \ell' & L\\
  -1 & 1 & 0\\
  \end{pmatrix} 
  + 
  \left[-c^{v,E}_{LM}(f) X^L_{\ell \ell'} - i c^{v,B}_{LM}(f) (1-X^L_{\ell \ell'}) \right]
\begin{pmatrix}
  \ell & \ell' & L\\
  1 & 1 & -2\\
  \end{pmatrix} 
  \right\},
\end{align}
where see Eq.~(\ref{eq:z_lm_corr_v}) for the expression of $\braket{z^*_{f,\ell m} z_{f,\ell' m'}}$ due to spin-1 GWs.
From this, the ORF coefficients are
\begin{align}
  F^{L,v,X,EE}_{\ell \ell'} = \frac{E^v_\ell E^v_{\ell'}}{z^v_\ell z^v_{\ell'}} F^{L,v,X,zz}_{\ell \ell'},
\end{align}
and the right column in Tables~\ref{tab:F_tb} for $F^{L,v,X,EB}_{\ell \ell'}$, \ref{tab:F_bt} for $F^{L,v,X,BE}_{\ell \ell'}$, and \ref{tab:F_bb} for $F^{L,v,X,BB}_{\ell \ell'}$.
Figure~\ref{fig:g_v_I_eb} shows the result of $\Gamma^{v,I,EB}_{LM}$.
In the figure, we can see again $\Gamma^{v,I,EB}_{00} = 0$, similar to the spin-2 GW case (Fig.~\ref{fig:g_t_I_eb}).

\subsection{Cross-correlation between redshift response and deflection fields}
\label{subsec:cross}

We finally discuss the cross-correlation between the redshift response and the angular deflection. 
We here consider the two-point correlation function of $z$ and $\delta \vec n$:
\begin{align}
  \vev{z^*_{f}(\hat n_a) \delta n_{f,i}(\hat n_b)} &= \sum_{\ell m} \sum_{\ell' m'} \sum_{S}\vev{z^*_{f,\ell m} S_{f,\ell' m'}} Y^*_{\ell m}(\hat n_a) Y^S_{(\ell' m'),i}(\hat n_b) \nonumber \\
  &= \sum_{p=t,v} \sum_X \sum_{S} i^X \left[ I^p_0(f) \sum_{L,M} c^{p,X}_{LM}(f) \Gamma^{p,X,zS}_{LM,i}(\hat n_a, \hat n_b) \right],
  \label{eq:zeb_corr}\\
  \vev{\delta n^*_{f,i}(\hat n_a) z_{f}(\hat n_b)} &= \sum_{\ell m} \sum_{\ell' m'} \sum_{S}\vev{S^*_{f,\ell m} z_{f,\ell' m'}} Y^{S*}_{(\ell m),i}(\hat n_a) Y_{\ell' m'}(\hat n_b) \nonumber \\
  &= \sum_{p=t,v} \sum_X \sum_{S} i^X \left[ I^p_0(f) \sum_{L,M} c^{p,X}_{LM}(f) \Gamma^{p,X,Sz}_{LM,i}(\hat n_a, \hat n_b) \right],
  \label{eq:ebz_corr}
\end{align}
where $S \in \{E,B\}$. 
The ORFs are given by 
\begin{tcolorbox}[ams align]
  \Gamma^{p,X,zS}_{(LM)i}(\hat n_a, \hat n_b) = (-1)^L \sqrt{\pi} \sum^{\ell_\tmax}_{\ell=\ell_\tmin} \sum^{\ell_\tmax}_{\ell' = \ell_\tmin} F^{L,p,X,zS}_{\ell \ell'}\left\{ Y_{\ell}(\hat n_a) \otimes Y^S_{(\ell')i}(\hat n_b)\right\}_{LM},
  \label{eq:orf_zeb} \\
  \Gamma^{p,X,Sz}_{(LM)i}(\hat n_a, \hat n_b) = (-1)^L \sqrt{\pi} \sum^{\ell_\tmax}_{\ell=\ell_\tmin} \sum^{\ell_\tmax}_{\ell' = \ell_\tmin} F^{L,p,X,Sz}_{\ell \ell'}\left\{ Y^S_{(\ell)i}(\hat n_a) \otimes Y_{\ell'}(\hat n_b)\right\}_{LM},
  \label{eq:orf_ebz}
\end{tcolorbox}  
where 
\begin{align}
    \left\{ Y_{\ell}(\hat n_a) \otimes Y^S_{(\ell')i}(\hat n_b)\right\}_{LM} = \sum^{\ell}_{m=-\ell} \sum^{\ell'}_{m'=-\ell'}\Braket{\ell m \ell' m' |L M} Y_{\ell m}(\hat n_a)Y^S_{(\ell' m')i}(\hat n_b), \\
  \left\{ Y^S_{(\ell)i}(\hat n_a) \otimes Y_{\ell'}(\hat n_b)\right\}_{LM} = \sum^{\ell}_{m=-\ell} \sum^{\ell'}_{m'=-\ell'}\Braket{\ell m \ell' m' |L M} Y^S_{(\ell m)i}(\hat n_a)Y_{\ell' m'}(\hat n_b).
\end{align}

\begin{figure}
        \centering \includegraphics[width=0.9\columnwidth]{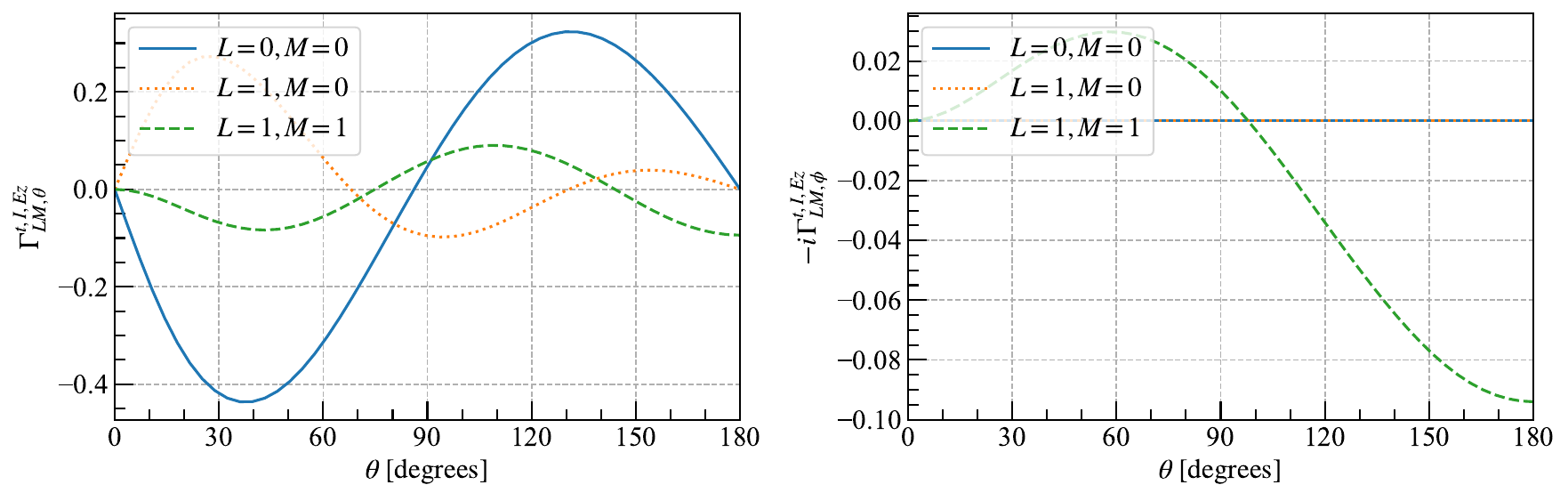}
        \caption{The ORFs of $\Gamma^{t,I,Ez}_{LM}(\theta)$ with $\ell_\tmax = 10$.
    }
        \label{fig:g_t_I_ez}
\end{figure}

\begin{figure}
        \centering \includegraphics[width=0.9\columnwidth]{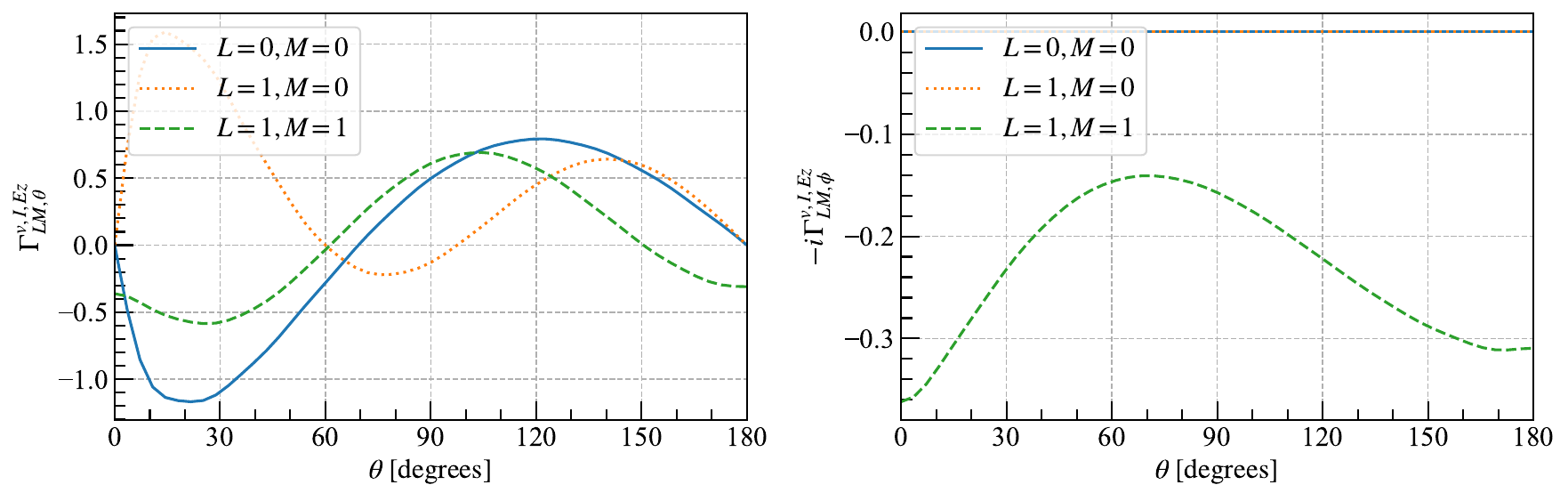}
        \caption{The ORFs of $\Gamma^{v,I,Ez}_{LM}(\theta)$ with $\ell_\tmax = 20$.
    }
        \label{fig:g_v_I_ez}
\end{figure}

We here recall that the redshift responses Eqs.~(\ref{eq:z_lm_t2}) (for spin-2 GWs) and (\ref{eq:z_lm_v2}) (for spin-1 GWs) are the same as the E-mode angular deflection Eqs.~(\ref{eq:e_lm_t2}) and (\ref{eq:e_lm_v2}) with $z^p_\ell \to E^p_\ell$, respectively.
From this, we can obtain the following relations for the ORF coefficients:
\begin{align}
  \label{eq:coeff_cross}
  F^{L,p,X,zS}_{\ell \ell'} = \frac{z^p_\ell}{E^p_\ell} F^{L,p,X,ES}_{\ell \ell'}, \ F^{L,p,X,Sz}_{\ell \ell'} = \frac{z^p_{\ell'}}{E^p_{\ell'}} F^{L,p,X,SE}_{\ell \ell'}.
\end{align}
Figures~\ref{fig:g_t_I_ez} and \ref{fig:g_v_I_ez} show the $\theta$ dependence of $\Gamma^{t,I,Ez}_{LM}$ and $\Gamma^{v,I,Ez}_{LM}$, respectively.

\section{Conclusion}
\label{sec:conclusions}

Here we have shown how to calculate the most general two-point correlation functions (or ORFs) for PTA timing residuals and the two components of the astrometric deflection due to a SGWB characterized by an arbitrary intensity and linear/circular-polarization pattern on the sky.  Results were presented for both the spin-2 (tensor) GWs that appear in general relativity as well as spin-1 (vector) GWs that may appear in alternative-gravity theories.

We have checked that our results for PTA timing residuals for an anisotropic, unpolarized, spin-2 and spin-1 GW background agree with Ref.~\cite{Gair:2015hra} up to $L \leq 3$. (While we have verified that our results are consistent for $L \leq 3$, we have only plotted up to $L\leq1$.)  For an anisotropic, polarized (either circular or linear) GW background, our results agree with those in Ref.~\cite{Bernardo:2023jhs}.  For astrometric angular deflections with $L = M = 0$ with spin-2 and spin-1 GW background, our results correspond with those in Refs.~\cite{Book:2010pf,Caliskan:2023cqm,Mihaylov:2018uqm,OBeirne:2018slh,Qin:2018yhy}.
The results presented here for astrometry with $L\geq1$ are new.
Note that we show only a limited number of plots in this paper and refer to \href{https://github.com/KeisukeInomata0/pyORFs}{our Github page} (Jupyter Notebook with Python code) for the other ORFs.

These computations are a first step that can be followed in subsequent work on detection forecasts, mission concept studies, and observational strategies.  We also leave the construction of estimators from astrometric deflections for anisotropy/polarization to future work, although they should follow analogous prior work.  And finally, we have restricted the calculation of the $F^L_{\ell\ell'}$ coefficients in the BiPoSH expansions of the ORFs to GWs that propagate at the speed of light.  It may be interesting to generalize in future work to nonluminal modes.

\section*{Acknowledgements}

This work was supported at Johns Hopkins by NSF Grant No.\ 2112699, the Simons Foundation, and the Templeton Foundation.
K.I. acknowledges the support of JSPS Postdoctoral Fellowships for Research Abroad.
C.T. acknowledges the support of the Department of Education Award \#P382G170104 (TITLE-VII). S.R.T. acknowledges support from NSF AST-2007993, NSF AST-2307719, and an NSF CAREER \#2146016. S.R.T. is a member of the NANOGrav collaboration, which receives support from NSF Physics Frontiers Center award number 1430284 and 2020265.
Any opinions, findings, interpretations, conclusions, or recommendations expressed in this material are those of its authors and do not represent the views of the Department of Education.

\appendix

\section{Wigner $D$-matrix}
\label{app:wigner}

In this Appendix, we summarize the properties of the Wigner $D$-matrix.
We consider the rotation of the coordinates:
\begin{align}
  \hat n^i \rightarrow {\hat n}^{\prime\, i} = R^{ij} \hat n_j,
\end{align}
where $R^{ij}$ is the rotation matrix given by~\cite{Khersonskii:1988krb} 
\begin{align}
R = 
  \begin{pmatrix}
  \cos \alpha \cos \beta \cos \gamma - \sin \alpha \sin \gamma & -\cos \alpha \cos \beta \sin \gamma - \sin \alpha \cos \gamma & \cos \alpha \sin \beta\\
  \sin \alpha \cos \beta \cos \gamma + \cos \alpha \sin \gamma & -\sin \alpha \cos \beta \sin \gamma + \cos \alpha \cos \gamma & \sin \alpha \sin \beta\\
  -\sin \beta \cos \gamma & \sin \beta \sin \gamma & \cos \beta \\
  \end{pmatrix},
\end{align}
where $\alpha$, $\beta$, and $\gamma$ are the Euler angles in the $z\text{-}y\text{-}z$ notation.
In this transformation, the spherical harmonics in the two coordinates are related to the Wigner $D$-matrix as~\cite{Goldberg:1966uu}
\begin{align}
  Y_{\ell m}(\hat n^\prime) = \sum_{m'} Y_{\ell m'}(\hat n) D^{(\ell)}_{m' m}(R^{-1}),
  \label{eq:y_dy0}
\end{align}
where $R^{-1}$ is the inverse rotation of $R$.
We here express Wigner $D$-matrix as\footnote{In Ref.~\cite{Goldberg:1966uu}, Wigner $D$-matrix is expressed as $D^{(\ell)}_{m' m}(R^{-1}) = D^{(\ell)}_{m' m}(\alpha\beta\gamma)$.}
\begin{align}
  D^{(\ell)}_{m m'}(R) = D^{(\ell)}_{m m'}(\alpha, \beta, \gamma),\ \   D^{(\ell)}_{m'm}(R^{-1}) = D^{(\ell)}_{m'm}(-\gamma, -\beta, -\alpha).
\end{align}
Also, its complex conjugate is given by~\cite{Khersonskii:1988krb}
\begin{align}
  D^{(\ell)*}_{m m'}(R) =  (-1)^{m'- m}D^{(\ell)}_{-m -m'}(R).
\end{align}
The spin-weighted spherical harmonic is related to Wigner $D$-matrix as~\cite{Okamoto:2002ik,Shiraishi:2010sm,Gair:2015hra}\footnote{The normalization of the spin-weighted spherical harmonics is slightly different from that in Ref.~\cite{Goldberg:1966uu}. With the normalization of Ref.~\cite{Goldberg:1966uu}, we find $_s Y_{\ell m}(\theta,\phi) = (-1)^{m-s} \sqrt{\frac{2\ell + 1}{4\pi}} D^{(\ell)}_{-m s}(\phi, \theta, \gamma) \ee^{i s \gamma}$. }
\begin{align}
  \,_s Y_{\ell m}(\theta,\phi) = (-1)^m \sqrt{\frac{2\ell + 1}{4\pi}} D^{(\ell)}_{-m s}(\phi, \theta, \gamma) \ee^{i s \gamma}.
\end{align}
Wigner D-matrix can be expressed with Wigner small $d$-matrix $d^\ell_{m m'}(\beta)$,
\begin{align}
  D^{(\ell)}_{m m'}(\alpha, \beta, \gamma) = \ee^{-i m\alpha} d^\ell_{m m'}(\beta) \ee^{-im' \gamma}.
\end{align}
Using $d^{\ell*}_{m m'}(\beta) = d^\ell_{m' m}(-\beta)$, we can see that Wigner $D$-matrix satisfies the following relation:
\begin{align}
   D^{(\ell)}_{m' m}(R^{-1}) = D^{(\ell)*}_{m m'}(R).
\end{align}
Using this, we can reexpress Eq.~(\ref{eq:y_dy0}) as 
\begin{align}
  Y_{\ell m}(\hat n^\prime) = \sum_{m'} D^{(\ell)*}_{m m'}(\alpha, \beta, \gamma) Y_{\ell m'}(\hat n).
  \label{eq:y_dy}
\end{align}

In the main text (e.g. Eq.~(\ref{eq:z_ydz})), we consider the transformation from the coordinates where the $z$-axis is aligned with the GW propagation direction ($\hat k \parallel \hat z$) to general coordinates. 
Let us here relate the angular coefficients in the two coordinates by using the above expressions. 
In general, we can expand some function of $\hat n$ as 
\begin{align}
  f_{\hat k}(\hat n) = \sum_{\ell m} \tilde f_{\ell m}(\hat k) Y_{\ell m}(\hat n),
\end{align}
where the subscript $\hat k$ means that $\hat n$ is defined in the coordinates with $\hat k$ aligned along the $z$-axis. 
We now move to other coordinates where $\hat \Omega (\neq \hat k)$ is aligned along the $z$-axis. 
Even if the physical observation direction is fixed, the coordinate transformation changes the observation direction vector as $\hat n \rightarrow \hat n'$.
That is, we have 
\begin{align}
  f_{\hat k}(\hat n) = f_{\hat \Omega}(\hat n') = \sum_{\ell m} \tilde f_{\ell m}(\hat k) Y_{\ell m}(\hat n) = \sum_{\ell m} \tilde f_{\ell m}(\hat \Omega) Y_{\ell m}(\hat n').
  \label{eq:f_nk_nome}
\end{align}
We here relate $Y_{\ell m}(\hat n)$ and $Y_{\ell m}(\hat n')$ using Wigner $D$-matrix.
We here define $R$ as 
\begin{align}
   \hat \Omega^i = R^{ij}\hat k_j.
\end{align}
The Euler angles are $\alpha = \phi_\Omega$ and $\beta = \theta_\Omega$, where $\hat \Omega = (\sin \theta_\Omega \cos \phi_\Omega, \sin \theta_\Omega \sin \phi_\Omega, \cos \theta_\Omega)$ in the $\hat k \parallel \hat z$ coordinates.
The final Euler angle $\gamma$ is arbitrary and we take $\gamma = 0$ in this work, which corresponds to make the polarization of `$+$', `$\times$', `$x$', and `$y$' aligned with $\hat \theta$ and $\hat \phi$ directions~\cite{AnilKumar:2023kvt}.
Then, we obtain 
\begin{align}
  {\hat n}^{\prime\, i} = (R^{-1})^{ij} \hat n_j.
\end{align}
Using this and Eq.~(\ref{eq:y_dy}), we obtain 
\begin{align}
    Y_{\ell m}(\hat n') &= \sum_{m'} D^{(\ell)*}_{m m'}(R^{-1}) Y_{\ell m'}(\hat n) \nonumber \\
    &=  \sum_{m'}Y_{\ell m'}(\hat n)  D^{(\ell)}_{m' m}(R).
\end{align}
Substituting this into Eq.~(\ref{eq:f_nk_nome}), we finally obtain 
\begin{align}
  f_{\hat k}(\hat n) = \sum_{\ell} \sum_{m m'} Y_{\ell m}(\hat n)  D^{(\ell)}_{m m'}(\phi_\Omega, \theta_\Omega, 0) \tilde f_{\ell m'}(\hat \Omega).
\end{align}
This corresponds to the relation in Eq.~(\ref{eq:z_ydz}).

%%%%%%%%%%%%%%%%%%%%%%%%%%%%%%%%%%%
%%%%%%%%%%%%%%%%%%%%%%%%%%%%%%%%%%%
%%%%%%%%%%%%%%%%%%%%%%%%%%%%%%%%%%%
\small
\bibliography{orf_pta_astrometry}{}

\end{document}